\documentclass[journal]{IEEEtran}
\usepackage{amsmath,amsfonts}
\usepackage{array}
\usepackage{textcomp}
\usepackage{stfloats}
\usepackage{url}
\usepackage{verbatim}
\usepackage{graphicx}
\usepackage{enumitem}
\usepackage{url}
\usepackage{graphics}
\usepackage{tikz}
\usepackage{multirow}
\usepackage{graphicx}
\usepackage{tabularx}
\usepackage{listings}
\usepackage{enumitem}
\usepackage{makecell}
\usepackage{algpseudocode}
\usepackage{amsmath}
\usepackage{amssymb}
\usepackage[all]{nowidow}
\usepackage{graphicx}

\usepackage{mathtools}
\usepackage{multirow}
\usepackage{algpseudocode}
\usepackage{url}

\usepackage[utf8]{inputenc}
\usepackage{enumitem}
\usepackage{url}
\usepackage{tikz}
\usepackage{caption}
\usepackage{subcaption}
\usepackage{multirow}
\usepackage{enumitem}
\usepackage{tabularx}
\usepackage{multirow}
\usepackage{amsthm}
\usepackage{listings}
\usepackage{float}
\usepackage{comment}
\usepackage{hyperref}
\usepackage{amsmath}
\usepackage[ruled,lined,boxed,linesnumbered]{algorithm2e}
\usepackage{amsthm}
\usepackage{balance}

\usepackage{titlesec}

\usepackage{caption}

\newcommand{\lefttriangle}{\tikz[baseline=-0.5ex]{\fill (0.0,0.0) -- (-0.3cm,1.0ex) -- (-0.3cm,-1.0ex) -- cycle;}}

\newcounter{reviewercount}
\newcounter{commentcount}

\newcommand{\aeditor}%
  {\bigskip\noindent {\bf COMMENTS OF THE ASSOCIATE EDITOR}%
  \setcounter{commentcount}{0}\par 
}

\usepackage[utf8]{inputenc}
\usepackage{enumitem}
\usepackage{url}
\usepackage{multirow}
\usepackage{graphicx}
\usepackage{enumitem}
\usepackage{tabularx}
\usepackage{multirow}
\usepackage{amsthm}
\usepackage{listings}
\usepackage{float}
\usepackage{comment}
\usepackage[ruled,lined,boxed,linesnumbered]{algorithm2e}

\usepackage{soul}

\usepackage{amsthm}
\theoremstyle{plain}

\theoremstyle{definition}

\newtheoremstyle{claim}
  {\topsep}
  {\topsep}
  {}
  {}
  {\itshape}
  {}
  {.5em}
  {\thmname{#1}\thmnumber{ #2}\thmnote{ (#3)}}

\SetKwInput{KwOutput}{Output}
\SetKwInput{KwInput}{Input} 

\usepackage{flushend}

\graphicspath{ {FIGS/} }

\begin{document}

\title{DG-RePlAce: A Dataflow-Driven GPU-Accelerated Analytical Global Placement Framework for Machine Learning Accelerators}
\author{Andrew~B.~Kahng,~\IEEEmembership{Fellow,~IEEE}
        and~Zhiang~Wang,~\IEEEmembership{Student Member,~IEEE}}

\maketitle

\begin{abstract}
Global placement is a fundamental 
step in VLSI physical design.
The wide \textcolor{black}{use} of 2D processing
element (PE) arrays in machine learning accelerators poses new challenges \textcolor{black}{of} scalability and Quality of Results (QoR) for state-of-the-art academic global placers. 
In this work, we develop {\em DG-RePlAce},
a new and fast GPU-accelerated global placement framework built on top of the
OpenROAD infrastructure \cite{OpenROAD},
which exploits the inherent dataflow and datapath structures of machine learning accelerators.
\textcolor{black}{
Experimental results with a variety of machine learning accelerators
using a commercial 12nm enablement show that,}
compared with {\em RePlAce} ({\em DREAMPlace}), our approach achieves an average reduction
in routed wirelength by $10\%$ ($7\%$) and total negative slack (TNS) by $31\%$ ($34\%$),
with faster global placement and on-par total runtimes relative to {\em DREAMPlace}.
Empirical studies on the {\em TILOS MacroPlacement Benchmarks} \cite{MacroPlacement} 
further demonstrate that post-route improvements over {\em RePlAce} and
{\em DREAMPlace} may reach beyond the motivating application to machine 
learning accelerators.


\end{abstract}

\maketitle

\section{Introduction}
\label{sec:intro}

Global placement is a fundamental step in VLSI physical design \textcolor{black}{that} determines the locations of standard cells and macros in a layout.
The backend design closure flow requires a fast placement engine for rapid design prototyping, 
feeding back to \textcolor{black}{synthesis,} and guiding optimization.
However, emerging machine learning accelerators
have introduced new challenges for global placement.
On the one hand,  machine learning accelerators with millions of standard cells and macros
raise runtime concerns for the design closure process.
On the other hand, 
machine learning accelerators featuring 2D processing element (PE) arrays, such as
systolic arrays \cite{Quinton87}, have gained prominence because of their efficiency in convolutional neural network computations \cite{JouppiYPPA17}. 
The dataflow and datapath architectures of modern machine learning accelerators exhibit substantial differences compared to
\textcolor{black}{those} of traditional datapath designs,
\textcolor{black}{requiring} dedicated treatment during global placement to achieve decent \textcolor{black}{quality of results} (QoR).

To address \textcolor{black}{aspects of} the aforementioned challenges, 
several global placers have been proposed over the past decades.
\textcolor{black}{To improve runtime,  researchers have focused on parallelizing global placement algorithms to
leverage the computational substrates provided by multi-core CPUs and GPUs.}
\cite{GesslerBS20} introduces a multi-threaded shared-memory implementation of RePlAce \cite{ChengKKW19} using off-the-shelf multi-core CPU hardware.
\cite{CongZ09} and \cite{LinW18}
propose GPU-accelerated analytical placers by parallelizing the computation of the 
Logarithm-Sum-Exponential (LSE) wirelength function
as well as \textcolor{black}{the} density function.
Recently, {\em DREAMPlace} \cite{LinJGLD20, LiaoLCLLY22}
and {\em Xplace} \cite{LiuFWY22}
\textcolor{black}{have implemented} the approach of {\em RePlAce} on GPU
by casting the placement problem as a neural network training 
\textcolor{black}{problem; these works} demonstrate the superiority
of GPU-accelerated global placers.
While {\em DREAMPlace} has already achieved 
significant runtime improvement relative to {\em RePlAce}, our aim is to push these boundaries \textcolor{black}{even} further by leveraging optimized data structures and a new parallel wirelength gradient computation algorithm.

\textcolor{black}{Netlist and register clustering are widely used in placement to achieve better QoR. 
\cite{LuPL21}, \cite{LuYLR22} use clustering information 
to induce soft placement constraints
for commercial placers.
\cite{ChangLJN19} proposes a register clustering approach to balance
clock power reduction and timing degradation.
\cite{LiuJLDZW22} adopts c-spectral clustering to reduce the problem size and ensure that standard-cell clusters and macros are of comparable size.
\cite{KahngKKMPP24} proposes a PPA-aware clustering approach
that takes into account timing, power and logical hierarchy during netlist clustering, effectively 
accelerating global placement runtime and improving post-route
QoR.
Other pioneering works adopt the clustering idea to exploit the dataflow and datapath structures during macro placement and global placement.} 
\cite{Vidal-ObiolsCPGM21} and \cite{KahngVW22}
exploit RTL information and dataflow to guide macro placement.
\cite{LinDYCL21} integrates the dataflow information into the mixed analytical global placement framework through virtual objects.
Furthermore, \cite{FangZHLY22} introduces the first global placement framework that exploits the datapath regularity of 2D PE arrays.  
In \textcolor{black}{our present} work, we propose a new, fast GPU-accelerated global placement framework, exploiting both dataflow information and datapath regularity.
Our approach ultimately guides global placement towards better
\textcolor{black}{QoR}.
The main contributions of this paper are as follows.
\begin{itemize}
    \item We propose {\em DG-RePlAce}, a new and fast global placer that leverages the intrinsic dataflow and datapath structures of machine learning accelerators,
    to achieve high-quality global placement.

    \item
    {\em DG-RePlAce} is built on top of the OpenROAD infrastructure
    \textcolor{black}{with a permissive open-source license}, enabling other
    researchers to readily adapt it for other enhancements.\footnote{The source
    code is available in the {\em DG-RePlAce} GitHub repository \cite{DG-RePlAce}.}
    
    \item We propose efficient data structures and algorithms to further speed up the global placement.  Experimental results on a variety of machine learning accelerators show that, our approach is respectively on average $22.49X$ and $1.75X$ faster
    than {\em RePlAce} and {\em DREAMPlace} in terms of global placement runtime. Overall
    turnaround time is on par with that of {\em DREAMPlace}, despite (one-time) file 
    IO runtime overheads that are due to OpenDB/OpenROAD integration.

   \item \textcolor{black}{Experimental} results \textcolor{black}{on a variety of}  machine learning accelerators \textcolor{black}{also}
    show that \textcolor{black}{in comparison with} {\em RePlAce} and {\em DREAMPlace}, our approach achieves an average reduction \textcolor{black}{of} routed wirelength by $10\%$ and $7\%$, and \textcolor{black}{of} total negative slack by $31\%$ and $34\%$, respectively.

   \item Experimental results on the two largest {\em TILOS MacroPlacement Benchmarks} 
   \cite{MacroPlacement} testcases show that compared with {\em RePlAce} and {\em DREAMPlace},
   {\em DG-RePlAce} achieves much better timing metrics (WNS and TNS) measured post-route optimization. This suggests that the proposed dataflow-driven methodology is not limited to machine learning accelerators. 

\end{itemize}

The remaining sections are organized as follows. 
Section \ref{sec:prelim} introduces the terminology and background.
Section \ref{sec:approach} discusses our approach.
Section \ref{sec:exp} shows experimental results,
and Section \ref{sec:conclusion} concludes the paper and
outlines future research directions.

\section{Preliminaries}
\label{sec:prelim}

In this section, we begin by discussing the fundamentals of the systolic array structure in Section \ref{sec:systolic_array_structure}. Following this, we delve into the electrostatics-based placement formulation,
\textcolor{black}{which is incorporated into DG-RePlAce,
in Section \ref{sec:electrostatics}.}
Finally, we examine previous works on dataflow-driven placement in Section \ref{sec:dataflow-macro-placement}.
\textcolor{black}{
Table \ref{tab:terms} summarizes important 
terms and their meanings; for clarity, 
we give 1-dimensional ($x$ component) notation.}


\begin{table}[!htb]
  \centering
  \caption{Terminology and notation.}
  \resizebox{0.8\columnwidth}{!}{%
  \begin{tabular}{|l|l|}
  \hline
  Notation & Description  \\ \hline \hline
  $v$ & instance (standard cell or macro) \\ \hline
  $p$ & instance pin or input-output pin \\ \hline
  $e$ & net $e = \{ p \}$ \\ \hline 
  $V$ & Set of \textcolor{black}{all} instances \textcolor{black}{\{$v$\}} \\ \hline 
  \textcolor{black}{$E$} & Set of \textcolor{black}{all} nets \textcolor{black}{(hyperedges)} \textcolor{black}{\{$e$\}} \\ \hline
  $P$ & Set of \textcolor{black}{all} pins \textcolor{black}{\{$p$\}} \\ \hline
  $WL_{grad_x}(p)$ & Wirelength gradient on pin $p$ \\ \hline
  $x_p$ & $x$ coordinate of pin $p$ \\ \hline
  $x_e^+$ & $max_{i \in e}$ $x_i$, $\forall e \in E$ \\ \hline 
  $x_e^-$ & $min_{i \in e}$ $x_i$, $\forall e \in E$ \\ \hline
  $a_i^+$ & \textcolor{black}{$exp({\frac{x_i - x_e^+}{\gamma}})$}, $\forall i \in e$, $e \in E$ \\ \hline 
  $a_i^-$ & \textcolor{black}{$exp({- \frac{x_i - x_e^-}{\gamma}})$}, $\forall i \in e$, $e \in E$  \\ \hline
  $b_e^+$ & $\sum_{i \in e}{a_i^+}$, $e \in E$ \\ \hline 
  $b_e^-$ & $\sum_{i \in e}{a_i^-}$, $e \in E$ \\ \hline
  $c_e^+$ & $\sum_{i \in e}{x_i a_i^+}$, $e \in E$  \\ \hline
  $c_e^-$ & $\sum_{i \in e}{x_i a_i^-}$, $e \in E$ \\ \hline
  \textcolor{black}{$X_v$} & Instance location, $\forall v \in V$  \\ \hline
  $F_{WL_x}(v)$ & Wirelength force on instance $v$ \\ \hline
  PU & Processing unit \\ \hline
  PE & Processing element \\ \hline
  \end{tabular}
  }
  \label{tab:terms}
\end{table}

\subsection{Systolic Array Structure}
\label{sec:systolic_array_structure}

\begin{figure}[!htb]
    \centering
    \includegraphics[width=0.99\columnwidth]{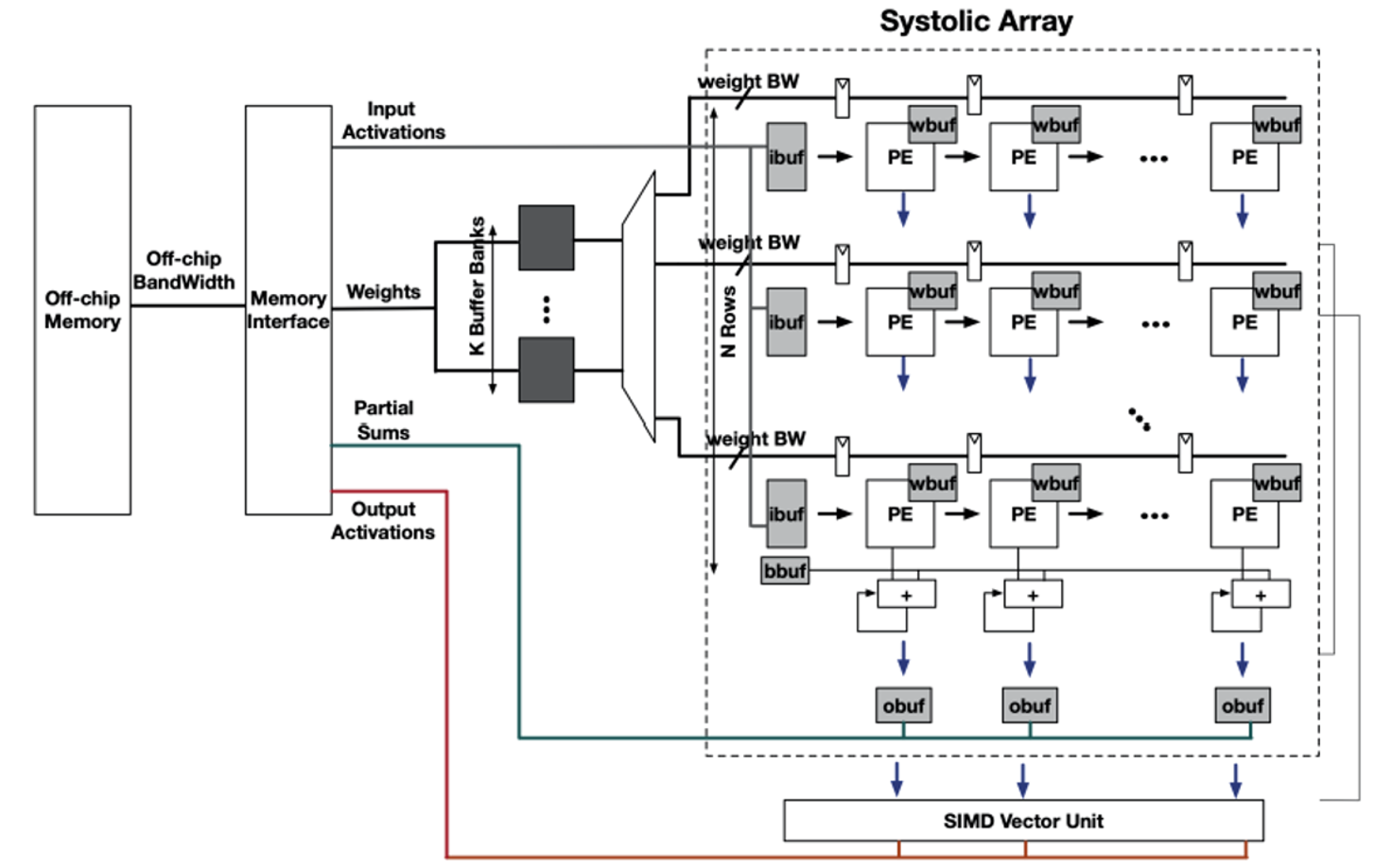}
    \caption{Illustrative execution flow of a systolic array-based machine learning accelerator 
    \textcolor{black}{(figure reproduced from \cite{EsmaeilzadehGGGK21})}.}
    \label{fig:systolic-array}    
\end{figure}

A systolic array \cite{EsmaeilzadehGGGK21} is a 2D array of $M \times N$ \textcolor{black}{{\em processing elements}} (PEs), which performs massively parallel convolution and matrix multiplication operations. 
A PE is often composed of a MAC (Multiply-Accumulate) unit and registers,
and PEs \textcolor{black}{that are} aligned in the same row collectively form a \textcolor{black}{{\em processing unit}} (PU).
Figure \ref{fig:systolic-array} shows an example execution flow of 
a systolic array-based machine learning accelerator.
Here, the input data horizontally propagate through the PEs,
\textcolor{black}{are} multiplied by the weights in each PE,
and then are accumulated vertically along the columns of the systolic array.
This structure restricts data transfers (multiple bitwidth) to neighboring PEs,
\textcolor{black}{thus} achieving  better performance and efficiency.

\subsection{Electrostatics-Based Placement}
\label{sec:electrostatics}

State-of-the-art academic global placers, such as {\em RePlAce}~\cite{ChengKKW19}  
and {\em DREAMPlace} \cite{LinJGLD20, LiaoLCLLY22}, usually adopt the 
electrostatics-based placement approach \textcolor{black}{\cite{LuCCLH15}}. 
Let \textcolor{black}{$(X_v, Y_v)^T$} \textcolor{black}{denote} the vectors of $x$-$y$ coordinates of movable instances.
The electrostatics-based placers formulate the global placement problem as follows:
\begin{equation}
  \textcolor{black}{
  \min_{X_v, Y_v} \sum_{e \in E}{WL(e ; X_v, Y_v)} +   \lambda \times D(X_v, Y_v)
  }
\end{equation}
where $WL(\cdot ;\cdot)$ is the wirelength cost function, 
$D(\cdot)$ is the instance density cost function 
and $\lambda$ is the weighting factor.
In this work, we use the weighted-average wirelength (WA) as the wirelength cost function,
where the $x$-component of WA for net $e$ is given by
\begin{equation}
   WL_e = \frac{\sum_{i \in e}{x_i \cdot exp({\frac{x_i}{\gamma}})}}{\sum_{i \in e}{ exp({\frac{x_i}{\gamma}})}} 
   - \frac{\sum_{i \in e}{x_i \cdot exp({- \frac{x_i}{\gamma}})}}{\sum_{i \in e}{ exp({- \frac{x_i}{\gamma}})}} 
\end{equation}
where $\gamma$ is a parameter \textcolor{black}{that controls} the smoothness 
and accuracy of the approximation to the half-perimeter wirelength (HPWL).
With the notations in Table \ref{tab:terms}, the gradient of WA wirelength
to a pin location $x_i$ is given as follows:
\begin{equation}
\label{eq:wirelength_grad}
    \frac{\partial WL_e}{\partial x_i}  
    =  \frac{(1 + \frac{x_i}{\gamma}) b_e^+ - \frac{1}{\gamma}c_e^+}{(b_e^+)^2} \cdot a_i^+
     -  \frac{(1 - \frac{x_i}{\gamma}) b_e^- + \frac{1}{\gamma}c_e^-}{(b_e^-)^2} \cdot a_i^-
\end{equation}

\subsection{Dataflow-Driven Placement}
\label{sec:dataflow-macro-placement}

To obtain high-quality macro placement,
human designers usually rely on their
understanding of the dataflow of a design
to determine the relative locations of macros.
However, this manual process is very time-consuming,
often \textcolor{black}{taking} several days to weeks to complete.
To automate this process, 
Vidal-Obiols et al. \cite{Vidal-ObiolsCPGM21} and {\em Hier-RTLMP} \cite{KahngVW23}
introduce dataflow-driven multilevel macro placement approaches.
However, both approaches apply the Simulated Annealing \cite{KirkpatrickGV83} algorithm to determine the locations of macros, 
resulting in poor runtime scalability \cite{AutoDMP}.
Lin et al. \cite{LinHCWW21} \textcolor{black}{present} an analytical-based placement
algorithm to handle dataflow constraints in mixed-size circuits.
Their approach initially assigns larger weights to nets connecting to
datapath-oriented objects, and then gradually
shrinks the weights according to the status of placement utilization.\footnote{
\cite{LinHCWW21} has reported an excellent dataflow-driven analytical
mixed-size placer. Unfortunately,
no testcases or executables can be released by their group.}
However,  their method requires dataflow constraints from designers and 
cannot handle the unique datapath regularity in machine learning accelerators (see Section \ref{sec:dataflow_extraction}). 
In this work, we incorporate the physical hierarchy extraction approach
in {\em Hier-RTLMP} \cite{KahngVW23} into the global placement framework to capture the dataflow information during global placement.
Besides, we pay special attention to the datapath regularity in machine learning accelerators during placement.

\section{Our Approach}
\label{sec:approach}

\begin{figure}[!htb]
    \centering
    \includegraphics[width=0.99\columnwidth]{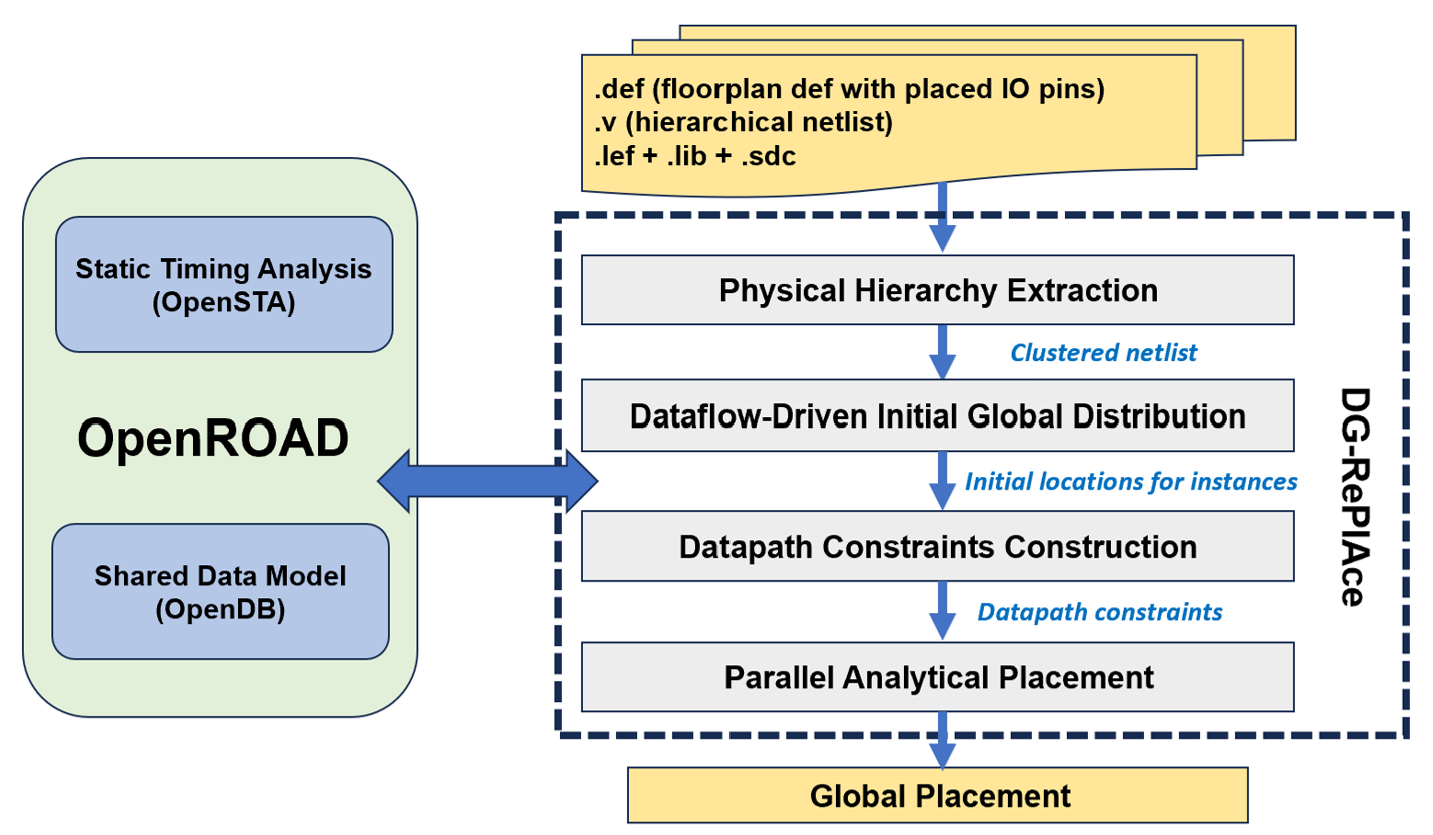}
    \caption{Overview of the proposed {\em DG-RePlAce} flow.}
    \label{fig:DG-RePlAce-flow}    
\end{figure}

The architecture of our {\em DG-RePlAce} framework is shown in Figure \ref{fig:DG-RePlAce-flow}.
The input is a synthesized hierarchical gate-level netlist and a floorplan .def
file that contains the block outline
and fixed IO pin or pad locations.
The output is a .def file with placed macros and standard cells.
{\em DG-RePlAce} is built on top of the open-source OpenROAD infrastructure \cite{OpenROAD}, \textcolor{black}{and} consists of four major steps.
\begin{itemize}
    \item \textbf{Physical Hierarchy Extraction} (Section \ref{sec:autocluster}): 
    During this \textcolor{black}{step}, we \textcolor{black}{convert} the structural netlist representation of the RTL design into a clustered netlist. 
    The instances within the same cluster are expected to remain close to each other
    during global placement.

    \item \textbf{Dataflow-Driven Initial Global Distribution} (Section \ref{sec:initial_placement}):
    During this \textcolor{black}{step}, we insert the dataflow information into the clustered netlist, and determine the location for each cluster through \textcolor{black}{our new} {\em GPU-accelerated parallel analytical placement} method.    
    Then, every instance within a cluster is positioned at the cluster's center\textcolor{black}{.}
    Furthermore, we incorporate pseudo net constraints \textcolor{black}{into} the original netlist, ensuring that instances belonging to the same cluster are placed in close proximity to \textcolor{black}{each other}.
    
    \item \textbf{Datapath Constraints Construction} (Section \ref{sec:dataflow_extraction}):
    \textcolor{black}{This step extracts} datapath information from the original netlist. Following this extraction, we
    transform the datapath information into pseudo net constraints.
    
    \item \textbf{Parallel Analytical Placement} (Section \ref{sec:global_placement}):
    At this \textcolor{black}{step}, we execute the GPU-accelerated mixed-size global placement on the original gate-level netlist, integrating the pseudo net constraints derived from the {\em Dataflow-Driven Initial Global Placement} and the {\em Datapath Constraints Construction} \textcolor{black}{steps}. 
    \textcolor{black}{Here,} we leverage parallel processing capabilities
    of GPU to accelerate the Nesterov's method in {\em RePlAce}.
    \textcolor{black}{In this work, we use the {\em GPU-accelerated parallel analytical placement} for both cluster-level and gate-level netlist placement.}
    
\end{itemize}

\noindent
\textcolor{black}{
In the remainder of this section, we will explain each step in detail. The runtime analysis for each step is presented in Figure \ref{fig:runtime} (see Section \ref{subsec:runtime}).}

\subsection{Physical Hierarchy Extraction}
\label{sec:autocluster}

During this \textcolor{black}{step}, we transform the original logical hierarchy 
into \textcolor{black}{a} physical hierarchy.
Much like the logical hierarchy, which is composed of logical modules, 
the physical hierarchy consists of physical clusters. 
In contrast to logical modules, a physical cluster consists of instances
that are expected to remain close to each other during global placement.
\textcolor{black}{Specifically, we employ the {\em Multilevel Autoclustering} component of the open-source {\em Hier-RTLMP} \cite{KahngVW23}}
to perform physical hierarchy extraction.\footnote{The detailed algorithm is presented \textcolor{black}{as Algorithm 2 in \cite{KahngVW23}. The source code is available in \cite{DG-RePlAce}.}}
Upon establishing the physical hierarchy, we convert the original gate-level netlist into a clustered netlist.

\subsection{Dataflow-Driven Initial Global Distribution}
\label{sec:initial_placement}

\textcolor{black}{
In this section, we first describe how to insert dataflow information into the clustered netlist. Then, we explain how we use the {\em GPU-accelerated parallel analytical placement} framework to distribute the clustered netlist evenly, and how we solve the divergence issue
by applying a bloat factor to each cluster. 
Finally, we discuss how we use the placed clustered netlist to guide the global placement process.
}

\begin{figure}[!h]
    \centering
    \includegraphics[width=0.85\columnwidth]{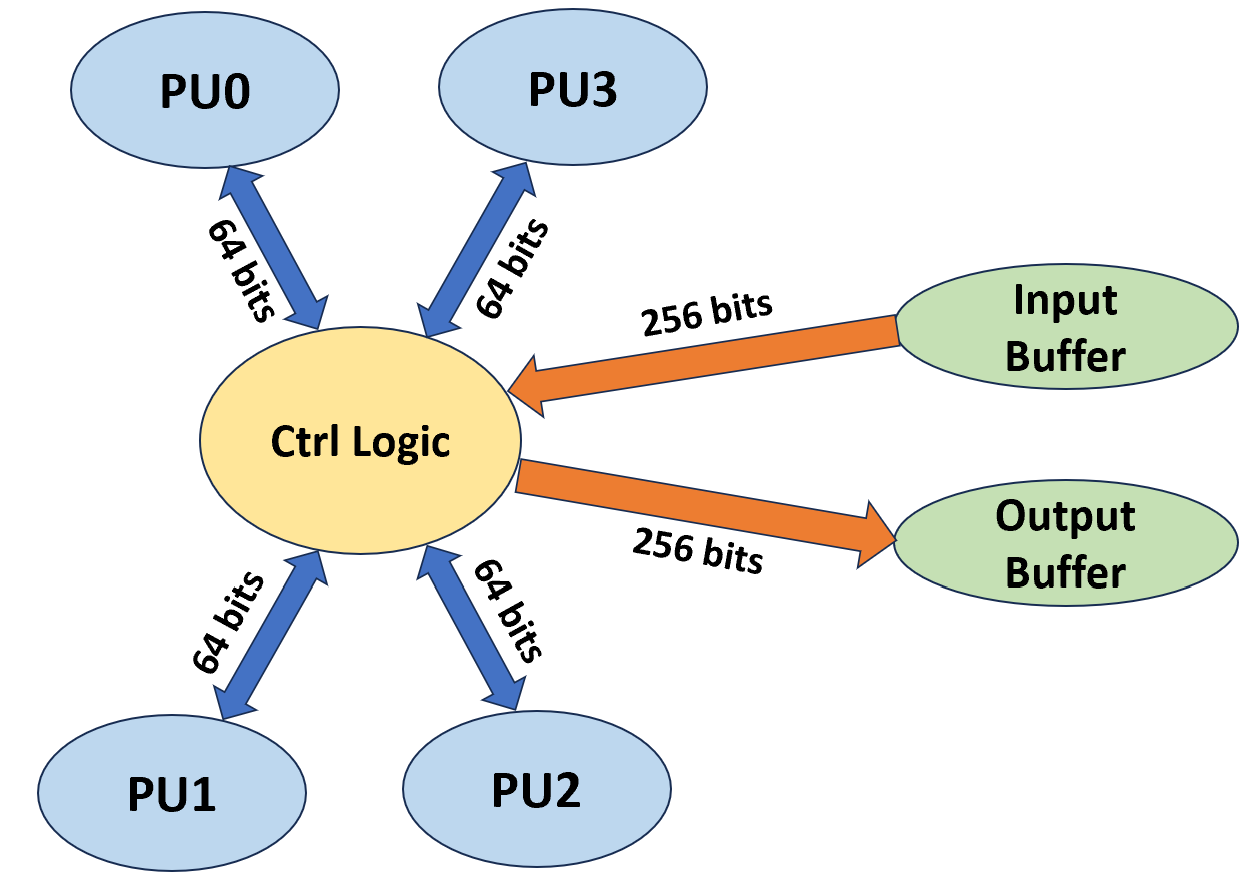}
    \caption{Dataflow visualization of the Tabla01 design \cite{EsmaeilzadehGGGK21}. }
    \label{fig:Tabla_dataflow}    
\end{figure}

After physical hierarchy extraction,
we have a clustered netlist in which the nodes
are clusters and the nets are bundled connections.
\textcolor{black}{We then} insert the dataflow information into the clustered netlist
through virtual connections.
\textcolor{black}{We use the term {\em dataflow} to refer to 
the way in which data moves between different functional
units of a netlist.}  The dataflow can be visualized as the high-level conceptual movements of data and how they are processed step by step.
Figure \ref{fig:Tabla_dataflow} shows the dataflow visualization of the Tabla01 design (see Section \ref{sec:exp} for details \textcolor{black}{of this and other testcases}).  
When backend engineers \textcolor{black}{perform} the place-and-route (P\&R) flow for a netlist, understanding the dataflow is critical for optimizing power, performance and area (PPA), as the dataflow determines how the netlist is pipelined and how the parallel processing is implemented.
We adopt the same idea as \cite{Vidal-ObiolsCPGM21,KahngVW22, KahngVW23} and transform the dataflow information into \textcolor{black}{{\em virtual connections}} between clusters.
The virtual connections ($virtual\_conn(A, B)$) between clusters A and B are defined as
\begin{equation}
    virtual\_conn(A, B) = \frac{info\_flow(A, B)}{2^{num\_hops}}
\end{equation}
Here, $info\_flow$ corresponds to connection bitwidth and 
$num\_hops$ is the length of \textcolor{black}{the} shortest path of registers between clusters.
\textcolor{black}{As pointed out in \cite{Vidal-ObiolsCPGM21}, 
the virtual connections between clusters capture 
the pipelined signal flow and the implementation of parallel processing.
This helps to ensure that the cluster placement aligns with the design's dataflow structure.}
When calculating the virtual connections between clusters,  we follow the same convention as \cite{KahngVW22, KahngVW23}.
If the register distance ($num\_hops$) between clusters is greater than 4, then no virtual connection is added.
In the example shown in Figure \ref{fig:Tabla_dataflow}, if the register distance between \emph{PU0} and \emph{Output Buffer} is 2, then the calculated virtual connections are 16, given that the connection bitwidth is 64 bits.

Upon incorporating the dataflow information into the clustered netlist, we call the {\em GPU-accelerated parallel analytical placement} framework, 
\textcolor{black}{for which details are given} in Section \ref{sec:global_placement},
to evenly distribute the clustered netlist.
However, directly working on the clustered netlist
could lead to divergence issues, \textcolor{black}{particularly if the layout has a large amount of whitespace.}
To solve the divergence issues, we introduce a \textcolor{black}{{\em bloat-shrink}} methodology guided by the final 
density overflow of the cluster placement (refer to Equation (37) in \cite{LuCCLH15} for the definition).
\textcolor{black}{In the bloat-shrink methodology, we initially bloat each cluster before cluster placement to achieve total cluster area that matches the area
of the placement region. If the cluster placement diverges, we then shrink each cluster to solve the divergence issue. More specifically, this} methodology includes two steps:
\begin{itemize}
    \item \textbf{Bloat:} We first bloat each cluster by applying a bloat factor ($bloat\_factor$), defined as     
    \begin{equation}
        bloat\_factor = \frac{\text{Area of placement region}}{\text{Total area of clusters}}
    \end{equation}

    \item  \textbf{Shrink:} 
    \textcolor{black}{If the cluster placement ends with a density overflow $cluster\_overflow$ that exceeds the target density overflow $target\_overflow$, we then shrink each cluster using a shrink factor ($shrink\_factor$).}
    $shrink\_factor$ is determined by dividing the target density overflow ($target\_overflow$) by the actual density overflow ($cluster\_overflow$):
    \begin{equation}
        shrink\_factor = \frac{target\_overflow}{cluster\_overflow}
    \end{equation}
    Here,  $target\_overflow$ ($target\_overflow$ = $0.2$ by default for cluster placement) is the convergence criterion for the Nesterov's approach.
\end{itemize}

\begin{figure}[!h]
    \centering
    \includegraphics[width=0.85\columnwidth]{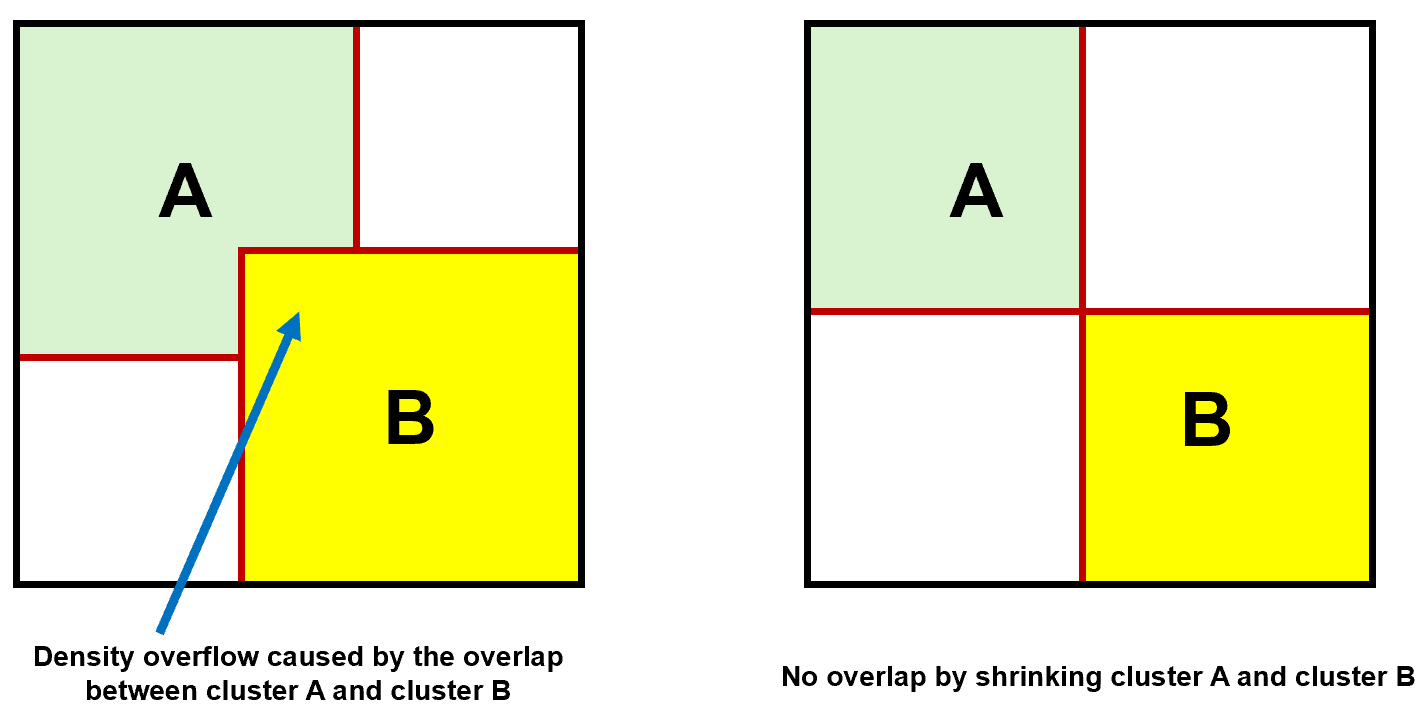}
    \caption{\textcolor{black}{Illustration of the bloat-shrink approach for reducing  $cluster\_overflow$. Left: density overflow caused by overlap between clusters 
    A and B; Right: removal of overlap by shrinking clusters A and B.}}
    \label{fig:bloat-shrink}    
\end{figure}

\textcolor{black}{
Global placers like {\em RePlAce} usually fill whitespace through filler insertion.
The fillers are equally sized rectangles, movable and disconnected (with zero pins). The additional density force created by the insertion of fillers helps squeeze the standard cells closer to their connected neighbors while still satisfying the density constraints \cite{LuCCLH15}.
In contrast to filler insertion used in {\em RePlAce} for filling whitespace, the bloat-shrink methodology ensures a feasible cluster placement solution
that satisfies the density overflow constraint.
The shrinking step is necessary because all clusters are square-shaped, which may lead to infeasible solutions.   
Figure \ref{fig:bloat-shrink} demonstrates how the bloat-shrink approach reduces $cluster\_overflow$.} 

After completing the placement of clusters, we place the instances
within each cluster at the cluster's center \textcolor{black}{to obtain}
a good initial placement.
Furthermore, \textcolor{black}{for each cluster, we add one pseudo net that
connects all of the instances within the cluster.} 
This ensures that instances belonging to the same cluster are placed in close proximity to \textcolor{black}{each other}.
However, we notice that these high-fanout pseudo-nets could cause convergence
problems. To address this issue, we transform the pseudo nets into 
multiple two-pin nets by the star model (i.e., by adding a pseudo vertex as the star's center, per cluster).
To ensure that the global placer follows the pseudo net constraints imposed by the clustering constraints, we initially assign a high penalty
factor $penalty\_factor_p$ to the pseudo nets, starting at the value of $penalty\_factor_{p0}$.
\textcolor{black}{With each successive iteration, the penalty factor}
is progressively decreased to allow for a more even distribution
of instances across the placement region.
The adjustment of the penalty factor $penalty\_factor_p$ is determined by the following equation:

    \begin{gather}
 \label{eq:init}   
         penalty\_factor_{p0} = exp(iter_0) \\
         penalty\_factor_p = \frac{penalty\_factor_{p0}}{exp(iter)}
    \end{gather}
where $iter$ is the current iteration number,
and $iter_0$ ($iter_0$ = 4 by default)  is used to determine the initial value.\footnote{\textcolor{black}{In our implementation, we set $penalty\_factor_p$ to 0 if $exp(iter)$ is NaN \cite{penalty_factor}.}}
To determine the default value of $iter_0$,
we study $iter_0$ values ranging from 0 to 9, 
utilizing Tabla01 and Tabla02 (see Table \ref{tab:benchmark}) as testcases.
The score for our evaluation is the routed wirelength, which we normalize
against the baseline results obtained from {\em RePlAce}.
Based on this experiment, we use $iter_0 = 4$ as the default.

\begin{figure*}[!t]
    \centering
    \begin{subfigure}[b]{0.50\textwidth}
        \centering
        \includegraphics[width=0.8\textwidth]{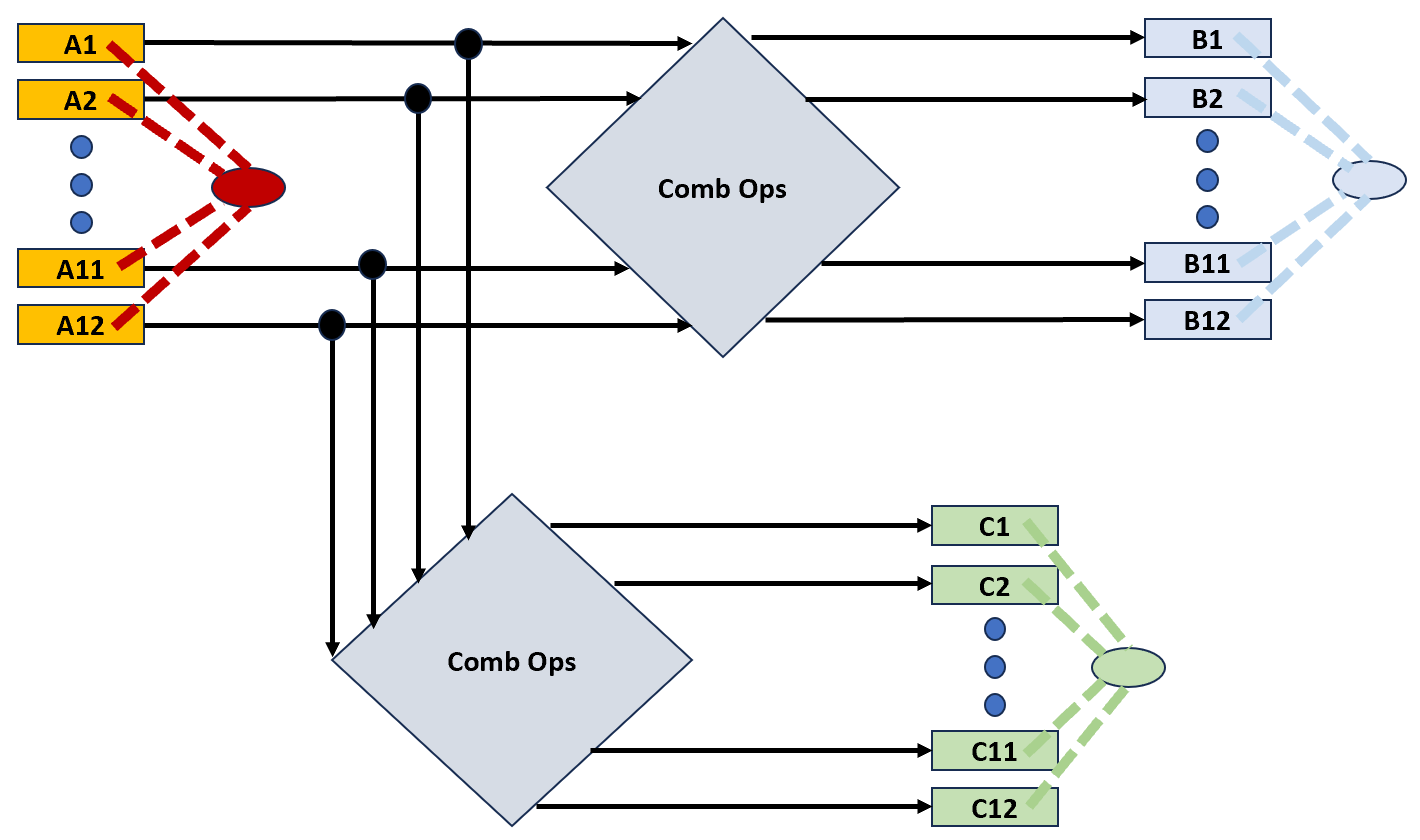}
        \caption{\textcolor{black}{Pseudo nets for bit stacks of a traditional datapath.}}
    \end{subfigure}
    \begin{subfigure}[b]{0.49\textwidth}
        \centering
        \includegraphics[width=0.95\textwidth]{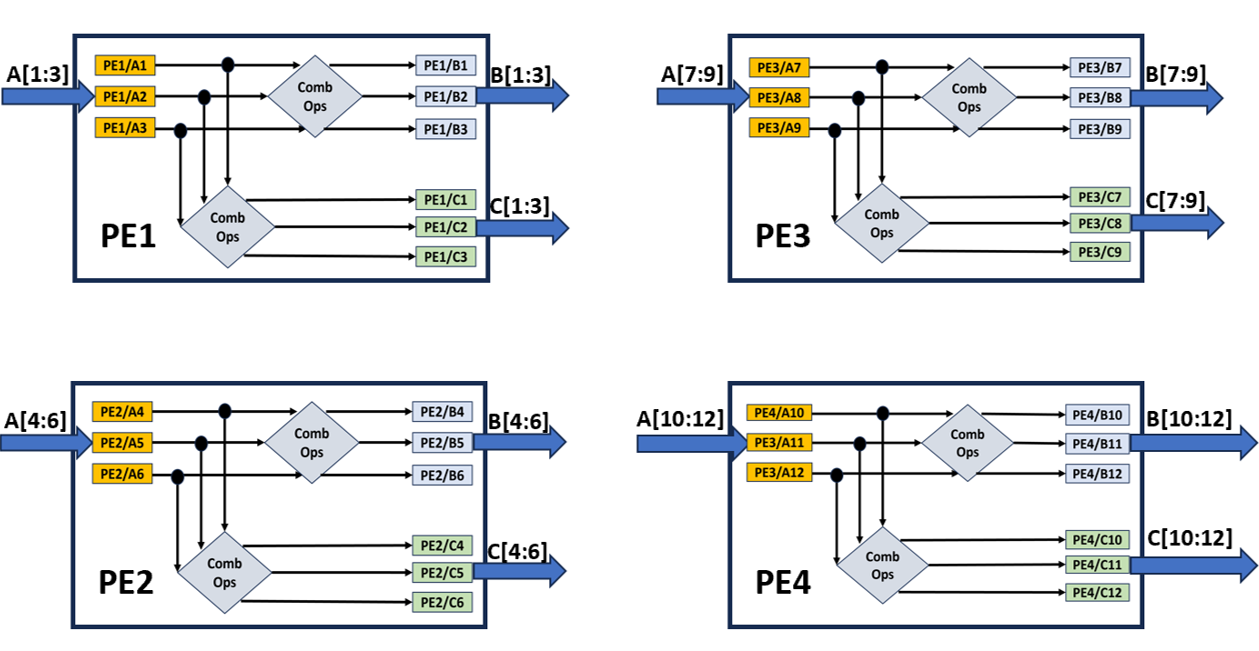}
        \caption{An example of \textcolor{black}{a} 2D PE array.}
    \end{subfigure}
    \begin{subfigure}[b]{0.50\textwidth}
        \centering
        \includegraphics[width=\textwidth]{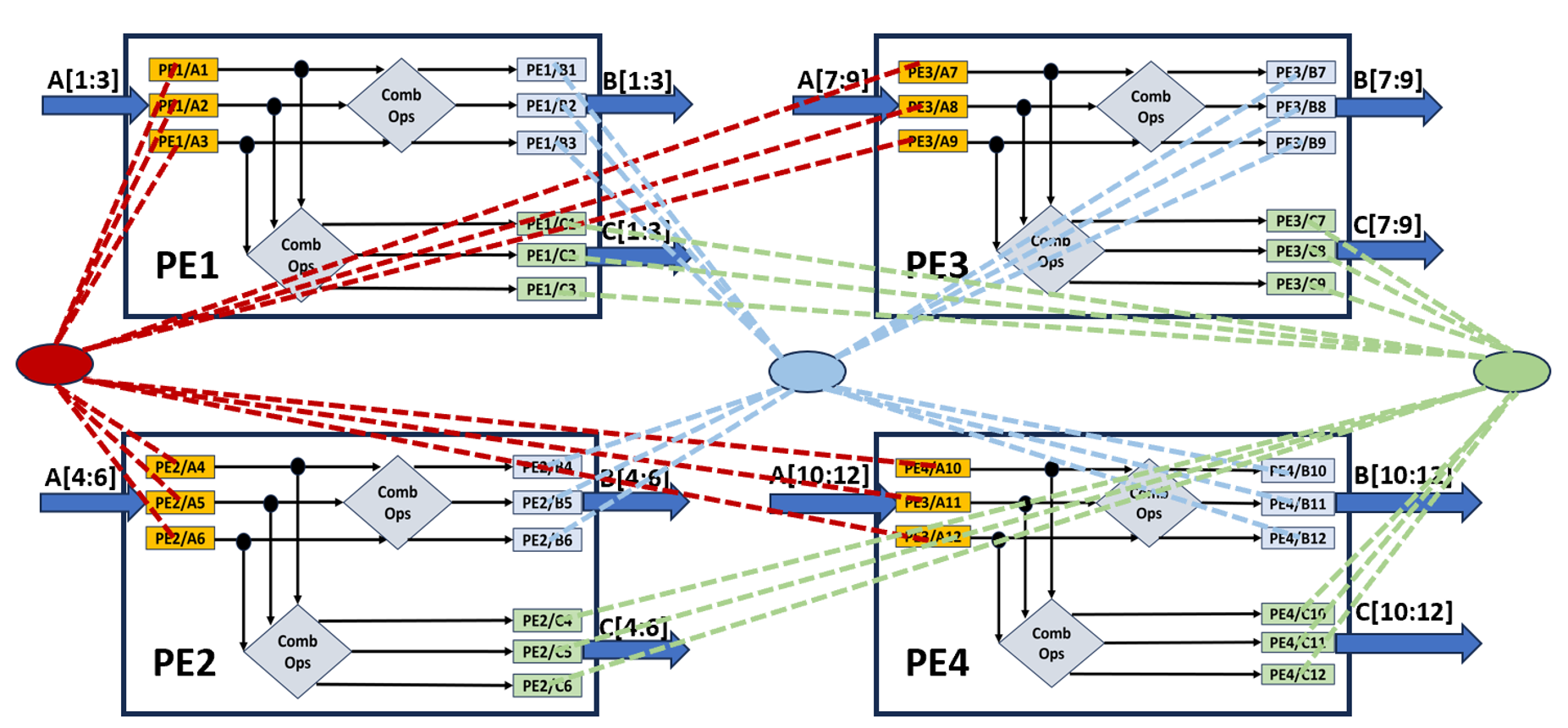}
        \caption{Applying pseudo nets on the 2D PE array.}
    \end{subfigure}
    \begin{subfigure}[b]{0.49\textwidth}
        \centering
        \includegraphics[width=0.97\textwidth]{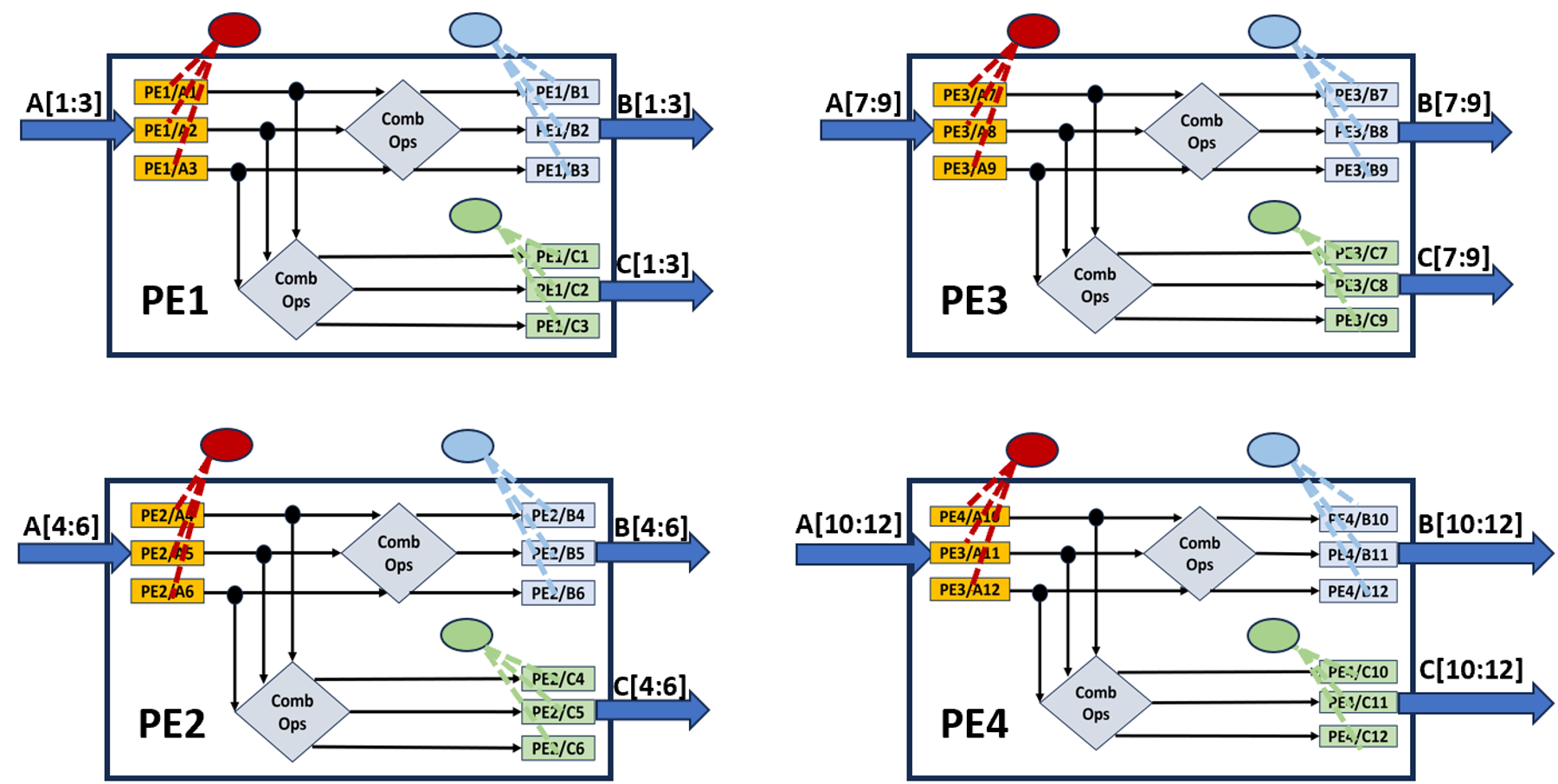}
        \caption{Datapath constraints on the 2D PE array.}
    \end{subfigure}
    \caption{Datapath constraints construction on the 2D PE array.}
    \label{fig:datapath_example}
\end{figure*}

\subsection{Datapath Constraints Construction}
\label{sec:dataflow_extraction}

After capturing the dataflow between clusters,
we examine the detailed data movement within
each cluster, i.e., datapath information.
\textcolor{black}{In contrast to dataflow,  
the {\em datapath} refers to the actual hardware components
and interconnections that implement the dataflow,
representing the paths that data traverse in a digital design.
More specifically,
the datapath is the circuit performing bit-wise data operations
in parallel on multiple bits \cite{ChouHC12}.}
Each operation \textcolor{black}{corresponds} to a dedicated functional block,
such as adder, register, buffer, multiplexer, multiplier, etc.
Fang et al.~\cite{FangZHLY22} further point out that
there is a significant difference between the datapath within
a systolic array and that of traditional datapath designs,
as shown in Figure \ref{fig:datapath_example}.

Figure \ref{fig:datapath_example}(a) shows the datapath in
traditional datapath designs, 
characterized by a continuous bit-sliced structure for operations
across different bits \cite{ChouHC12}.
In such scenarios, a pseudo net can be applied to each alignment group 
(i.e., A[1:12], B[1:12] and C[1:12]),
ensuring that the instances in each alignment group are placed in proximity.
In contrast, as depicted in Figure \ref{fig:datapath_example}(b),
the datapath in a systolic array is not continuous,
with operations for different bits across multiple PEs.
This may lead to overlaps of PEs when a pseudo net is 
directly applied to each 
alignment group, as shown in Figure \ref{fig:datapath_example}(c).
To address this issue, we propose to assign pseudo nets to local alignment
groups within each cluster. For example,
in Figure \ref{fig:datapath_example}(d), 
pseudo nets are independently applied to A[1:3] in PE1,
A[4:6] in PE2,
A[7:9] in PE3
and A[10:12] in PE4.
Here we also transform the pseudo nets into multiple two-pin nets according to the star model.
To maintain the integrity of local connectivity, 
we set the initial penalty factor $penalty\_factor_{p0}$ to 1 for the pseudo nets induced by datapath constraints.

\subsection{Parallel Analytical Placement}
\label{sec:global_placement}

\textcolor{black}{Our} GPU-accelerated mixed-size parallel 
analytical placement framework uses the same Nesterov's method as {\em RePlAce}, and is developed on top of the OpenROAD infrastructure. 
To minimize memory overhead, we integrate a data structure inspired
by Gessler et al. \cite{GesslerBS20}, which optimizes data locality for the
frequently accessed components during global placement.
\textcolor{black}{As pointed out by \cite{LinW18}},
the fast computation of wirelength gradient
and bin density is crucial for the efficiency of the global placer.
We adopt the parallel bin density computation algorithm 
from Gessler et al. \cite{GesslerBS20} (Algorithm 2 in \cite{GesslerBS20}).
For the fast computation of wirelength gradient, 
we introduce a novel parallel algorithm, presented in Algorithm \ref{alg:wirelength_grad}.
Our algorithm distinguishes itself from Algorithm 1 in {\em DREAMPlace} \cite{LinJGLD20} primarily in \textit{Lines 1-6} and \textit{Lines 12-18},
where we leverage net-level parallelization rather than pin-level parallelization to eliminate the need for atomic additions.
\textcolor{black}{Furthermore, it differs from Algorithm 2 in {\em DREAMPlace} \cite{LinJGLD20} 
primarily in \textit{Lines 7-11} and \textit{Lines 19-22},
where we implement pin-level computation parallelization with multiple threads
rather than the sequential computation within a single thread.
}
This approach is more efficient for managing high-fanout nets while maintaining comparable efficiency in handling low-fanout nets (see Section \ref{subsec:runtime} for details). 
Empirical results demonstrate that our algorithm is approximately
$3.25$X faster than the one implemented in {\em DREAMPlace} (Algorithm 2 in \cite{LinJGLD20}).

\begin{algorithm}[!h]
    \small
    \SetKwData{}{left}\SetKwData{This}{this}\SetKwData{Up}{Up}
    \SetKwInOut{Input}{input}\SetKwInOut{Output}{output}
    \KwInput{
    Instances $V$, 
    Nets $E$,
    Pins $P$
    and Instance locations $X_v$}
    \KwOutput{Wirelength force for each instance $F_{WL_x}(v)$ \\}
    \BlankLine{}
    \For{each thread $0 \leq t < |E|$} {
       Define $e$ as the net \textcolor{black}{corresponding} to thread $t$; \\
       $x_e^+ \leftarrow max_{p \in e} x_p$;  \hfill \lefttriangle{} $x_e^+$ is in the global memory \\
       $x_e^- \leftarrow min_{p \in e} x_p$;  \hfill \lefttriangle{} $x_e^-$ is in the global memory \\
       $b_e^\pm \leftarrow 0$;  $c_e^\pm \leftarrow 0$; \hfill \lefttriangle{} $b_e^\pm$, $c_e^\pm$ are in the global memory \\
    }   

    \For{each thread $0 \leq t < |P|$} {
       Define $p$ as the pin \textcolor{black}{corresponding} to thread $t$; \\
       Define $e$ as the net that pin $p$ belongs to; \\
       $a_p^\pm$ $\leftarrow$ $e^{\pm \frac{x_p - x_e^\pm}{\gamma}}$; 
       \hfill \lefttriangle{} $a_p^\pm$ is in the global memory \\
    }      

    \For{each thread $0 \leq t < |E|$} {
       Define $e$ as the net \textcolor{black}{corresponding} to thread $t$; \\
       \For{pin $p \in e$} {
          $b_e^\pm \leftarrow b_e^\pm + a_p^\pm$; \\
          $c_e^\pm \leftarrow c_e^\pm + x_p a_p^\pm$; \\
       }
    }      

    \For{each thread $0 \leq t < |P|$} {
       Define $p$ as the pin \textcolor{black}{corresponding} to thread $t$; \\
       Compute the wirelength gradient of pin $WL_{grad_x}(p)$ using \textcolor{black}{Equation (\ref{eq:wirelength_grad})};
       \hfill \lefttriangle{} $WL_{grad_x}(p)$ is in the global memory \\
    }      

    \For{each thread $0 \leq t < |V|$} {
       Define $v$ as the instance \textcolor{black}{corresponding} to thread $t$; \\
       $F_{WL_x}(v) \leftarrow 0.0$  \hfill \lefttriangle{} $F_{WL_x}(v)$ is in the global memory \\
       \For{pin $p$ of $v$} {
          $F_{WL_x}(v) -= WL_{grad_x}(p)$; \\
       }
    }      
    
    \Return $F_{WL_x}(v)$
    \caption{Parallel Wirelength Gradient Computation.}
    \label{alg:wirelength_grad}
\end{algorithm}

\section{Experimental results}
\label{sec:exp}

{\em DG-RePlAce} is implemented with approximately 14K lines of C++ (and CUDA) with a Tcl
command line interface on top of the OpenROAD infrastructure \cite{OpenROAD}.
We run all experiments on a Linux server with an Intel Xeon E5-2690 CPU (48 threads) \textcolor{black}{with 256 GB RAM}
and an NVIDIA TITAN V GPU.

To show the effectiveness of our global placer,
the following three scenarios are evaluated and compared.
\begin{itemize}
    \item {\em RePlAce}: Global placement is done by {\em RePlAce}, which is the default global placer in the OpenROAD project~\cite{OpenROAD}.
    \item {\em DREAMPlace}: Global placement is performed by the latest version of {\em DREAMPlace} \cite{DREAMPlace4.0.0},
    which is the state-of-the-art GPU-accelerated global placer.
    \textcolor{black}{The
    default hyperparameter settings that we use for 
    {\em DREAMPlace} are from \cite{DREAMPlace_params}.}
    \item {\em DG-RePlAce}: Results are obtained using our global placer.
\end{itemize}

Our experiments use the following flow.
(1) We first synthesize the design using a 
\textcolor{black}{state-of-the-art} commercial synthesis tool,
preserving the logical hierarchy.
(2) Next, we determine the core size of the testcase and place all the IO pins using a manually-developed script \textcolor{black}{(see \cite{VeriGOOD-ML})}.
(3) Then, the global placement is performed using different methods ({\em RePlAce}, {\em DREAMPlace} and {\em DG-RePlAce}).
(4) Finally, we use a state-of-the-art commercial P\&R tool, \textcolor{black}{Cadence Innovus 21.1,}
to finish the legalization of macros, detailed placement of
standard cells and routing. We follow the SP\&R scripts in the \textcolor{black}{public} {\em MacroPlacement repository} \cite{MacroPlacement}.
All metrics are collected after post-route optimization.
All studies use a commercial foundry 12nm technology 
(13 metal layers)
with cell library and memory generators from a leading IP provider.

\textcolor{black}{
In this section, we first present the results for two types of machine learning accelerators:
non-DNN machine learning accelerators (Tabla designs) and DNN machine learning accelerators (GeneSys designs), detailed in Section \ref{subsec:main_results}.
Then, we discuss the runtime comparison between {\em DG-RePlAce} and {\em DREAMPlace}
in Section \ref{subsec:runtime}.
Following this, Section \ref{sec:comp_hier_rtlmp} compares {\em DG-RePlAce} with the 
dataflow-driven macro placer {\em Hier-RTLMP}, which uses the same method 
to perform physical hierarchy extraction.
Next, we study the respective effects of dataflow and datapath constraints by conducting an 
ablation study of {\em DG-RePlAce}, in Section \ref{sec:ablation}.
\textcolor{black}{Last}, we apply {\em DG-RePlAce} to large non-machine learning testcases 
in Section \ref{sec:tilos_result}, 
demonstrating the versatility and potential
benefit of the proposed dataflow-driven approach beyond our motivating application
context of large-scale machine learning accelerators.
}

\subsection{Results on Machine Learning Accelerators}
\label{subsec:main_results}

We have validated our global placer using two types of machine learning accelerators (Tabla and GeneSys)
from the VeriGOOD-ML platform \cite{EsmaeilzadehGGGK21}.
The Tabla accelerators are designed for training and inference for non-DNN machine learning algorithms,
and the GeneSys accelerators are for DNN machine learning algorithms.
Both Tabla and Genesys adopt the systolic array structure, 
thus each design has an $m \times n$ PE array.
The major characteristics of the testcases are summarized in Table \ref{tab:benchmark}.

\begin{table}[!t]
  \caption{\small Benchmarks.
  ``Macro Util'' stands for macro utilization,
  which is defined as the total area of macros 
  divided by the core area.
  \textcolor{black}{``Util'' stands for utilization,
  which is defined as the total area of standard cells and macros with a 2$\mu$m halo width
  divided by the core area.}}
    \centering
    \begin{tabular}{|c|c|c|c|c|c|c|}
    \hline
    Designs & PE Array &  \# Macros & \# Std Cells & \# Nets & Macro Util & \textcolor{black}{Util} \\ \hline
    Tabla01 & 4 $\times$ 8 & 368 & 232K & 252K & 0.60 & \textcolor{black}{0.75} \\ \hline
    Tabla02 & 4 $\times$ 16 & 1232 & 441K & 486K & 0.59 & \textcolor{black}{0.79} \\ \hline
    Tabla03 & 8 $\times$ 8 & 760 & 372K & 408K & 0.58 & \textcolor{black}{0.78} \\ \hline
    Tabla04 & 8 $\times$ 16 & 2488 & 741K & 830K & 0.54 & \textcolor{black}{0.76} \\ \hline
    GeneSys01 & 16 $\times$ 16 & 368 & 986K & 1056K & 0.46 & \textcolor{black}{0.72} \\ \hline
    GeneSys02 & 16 $\times$ 16 & 368 & 1055K & 1135K & 0.52 & \textcolor{black}{0.71} \\ \hline
    \end{tabular}
    \label{tab:benchmark}
\end{table}

Table \ref{tab:result} shows the experimental results after completion
of post-route optimization. 
Rows represent testcases and global placement flows,
and columns give information on total routed wirelength,
power, worst negative slack (WNS), total negative slack (TNS),
runtime of global placement (GP) and turnaround time (TAT).\footnote{
The runtime of global placement refers to the runtime
required to distribute the original netlist across the placement region using the Nesterov's method. For {\em DREAMPlace}, 
we extract the relevant information from the following log file
entry: ``[INFO   ] DREAMPlace - non-linear placement takes xx seconds''.}
The metrics are normalized to protect foundry IP:
(i) wirelength and power are normalized to the {\em RePlAce} results,
and (ii) timing metrics (WNS and TNS) are normalized to the clock period which we leave unspecified.

\begin{table}[!t]
\caption{\textcolor{black}{Experimental results.
We highlight best values of metrics in blue bold font.
Data points \textcolor{black}{for WL, Power, WNS and TNS} are normalized.}
}
\label{tab:result}
\resizebox{1.00\columnwidth}{!} {
\centering
\begin{tabular}{|c|c|c|c|c|c|c|c|}
\hline 
\multicolumn{1}{|l|}{\makecell{Design}}             
    & \makecell{Global Placer}
    & \makecell{WL}   
    & \makecell{Power}
    & \makecell{WNS} 
    & \makecell{TNS}
    & \makecell{GP \\ ($s$) }
    & \makecell{TAT \\  ($s$)}
    \\ \hline \Xhline{2\arrayrulewidth}
\multirow{4}{*}{\makecell{Tabla01}}
& {\em RePlAce} & 1.00  & 1.00  & -0.180  & -81.349  & 150  & 204  \\  \cline{2-8}
& {\em DREAMPlace} & 0.98  & 1.01
                   & \textbf{\textcolor{blue}{-0.151}}  
                   & \textbf{\textcolor{blue}{-55.844}}  
                   & 14  
                   & \textbf{\textcolor{blue}{27}}  \\  \cline{2-8}
& {\em Hier-RTLMP} & 1.10   
                   & 1.00 
                   & -0.156  & -74.929  
                   & - & 2351  \\  \cline{2-8}
& {\em DG-RePlAce} & \textbf{\textcolor{blue}{0.93}}  
                   & \textbf{\textcolor{blue}{0.98}}  
                   & -0.180  & -62.622  
                   & \textbf{\textcolor{blue}{7}}  & 31  \\  \cline{2-8}
\hline \Xhline{2\arrayrulewidth}
\multirow{4}{*}{\makecell{Tabla02}}
& {\em RePlAce} & 1.00  & 1.00  & -0.197  & -22.695  & 374  & 476  \\  \cline{2-8}
& {\em DREAMPlace} & 0.95  
                   & \textbf{\textcolor{blue}{0.98}}  
                   & -0.188  & -24.188  & 24  
                   & \textbf{\textcolor{blue}{47}}  \\  \cline{2-8}
& {\em Hier-RTLMP} & 1.12
                   & 1.02
                   & \textbf{\textcolor{blue}{0.160}}  
                   & \textbf{\textcolor{blue}{-17.807}}  & -  
                   & 3613  \\  \cline{2-8}
& {\em DG-RePlAce} & \textbf{\textcolor{blue}{0.91}}  
                   & \textbf{\textcolor{blue}{0.98}}  
                   & -0.187
                   & -19.642  
                   & \textbf{\textcolor{blue}{13}}
                   & 50  \\  \cline{2-8}
\hline \Xhline{2\arrayrulewidth}
\multirow{4}{*}{\makecell{Tabla03}}
& {\em RePlAce} & 1.00  & 1.00  & -0.092  & -43.136  & 279  & 364  \\  \cline{2-8}
& {\em DREAMPlace} & 1.21  & 1.05  & -0.154  & -84.152  & 21  
                   & \textbf{\textcolor{blue}{41}}  \\  \cline{2-8}
& {\em Hier-RTLMP} & 0.98  & 0.99 & -0.136  & -88.578  & - 
                   & 3872  \\  \cline{2-8}                   
& {\em DG-RePlAce} & \textbf{\textcolor{blue}{0.88}}  
                   & \textbf{\textcolor{blue}{0.97}}  
                   & \textbf{\textcolor{blue}{-0.084}}  
                   & \textbf{\textcolor{blue}{-14.910}}  
                   & \textbf{\textcolor{blue}{16}}  
                   & 47  \\  \cline{2-8}
\hline \Xhline{2\arrayrulewidth}
\multirow{4}{*}{\makecell{Tabla04}}
& {\em RePlAce} & 1.00  & 1.02 
                & \textbf{\textcolor{blue}{-0.174}}  
                & \textbf{\textcolor{blue}{-48.756}}  & 689  & 883  \\  \cline{2-8}
& {\em DREAMPlace} & \textbf{\textcolor{blue}{0.83}}  
                   & \textbf{\textcolor{blue}{0.96}}  
                   & -0.177  & -62.990  & 47  
                   & \textbf{\textcolor{blue}{81}}  \\  \cline{2-8}
& {\em Hier-RTLMP} & 1.10  
                   & 1.06 
                   & -0.281  & -222.119  & -
                   & 8418  \\  \cline{2-8}                  
& {\em DG-RePlAce} & 0.85  & 0.97  & -0.219  & -54.755  
                   & \textbf{\textcolor{blue}{20}}  & 82  \\  \cline{2-8}
\hline \Xhline{2\arrayrulewidth}
\multirow{4}{*}{\makecell{GeneSys01}}
& {\em RePlAce} & 1.00  & 1.00  & -0.191  & -94.176  & 630  & 850  \\  \cline{2-8}
& {\em DREAMPlace} & \textbf{\textcolor{blue}{0.89}}  
                   & \textbf{\textcolor{blue}{0.98}}  & -0.213  & -101.598  & 61  & 101  \\  \cline{2-8}
& {\em Hier-RTLMP} & 0.97  
                   & 1.00
                   & \textbf{\textcolor{blue}{-0.110}}  & \textbf{\textcolor{blue}{-23.151}}  & -  & 3134  \\  \cline{2-8}
& {\em DG-RePlAce} & \textbf{\textcolor{blue}{0.89}}  
                   & \textbf{\textcolor{blue}{0.98}}  
                   & -0.162 
                   & -45.939
                   & \textbf{\textcolor{blue}{40}}  
                   & \textbf{\textcolor{blue}{100}}  \\  \cline{2-8}
\hline \Xhline{2\arrayrulewidth}
\multirow{4}{*}{\makecell{GeneSys02}}
& {\em RePlAce} & 1.00  & 1.00  & -0.132  & -14.937  & 752  & 972  \\  \cline{2-8}
& {\em DREAMPlace} & \textbf{\textcolor{blue}{0.89}}  
                   & \textbf{\textcolor{blue}{0.97}}  & -0.108  & -14.979  & 64  & 112  \\  \cline{2-8}
& {\em Hier-RTLMP} &  \multicolumn{6}{c|}{N/A} \\  \cline{2-8}
& {\em DG-RePlAce} & 0.94  & 0.99  
                   & \textbf{\textcolor{blue}{-0.062}}  
                   & \textbf{\textcolor{blue}{-7.78}}  
                   & \textbf{\textcolor{blue}{44}}  
                   & \textbf{\textcolor{blue}{111}}  \\  \cline{2-8}
\hline
\end{tabular}
}
\end{table}

\begin{figure}[!t]
    \centering
    \begin{subfigure}[b]{0.23\textwidth}
        \centering
        \includegraphics[width=\textwidth]{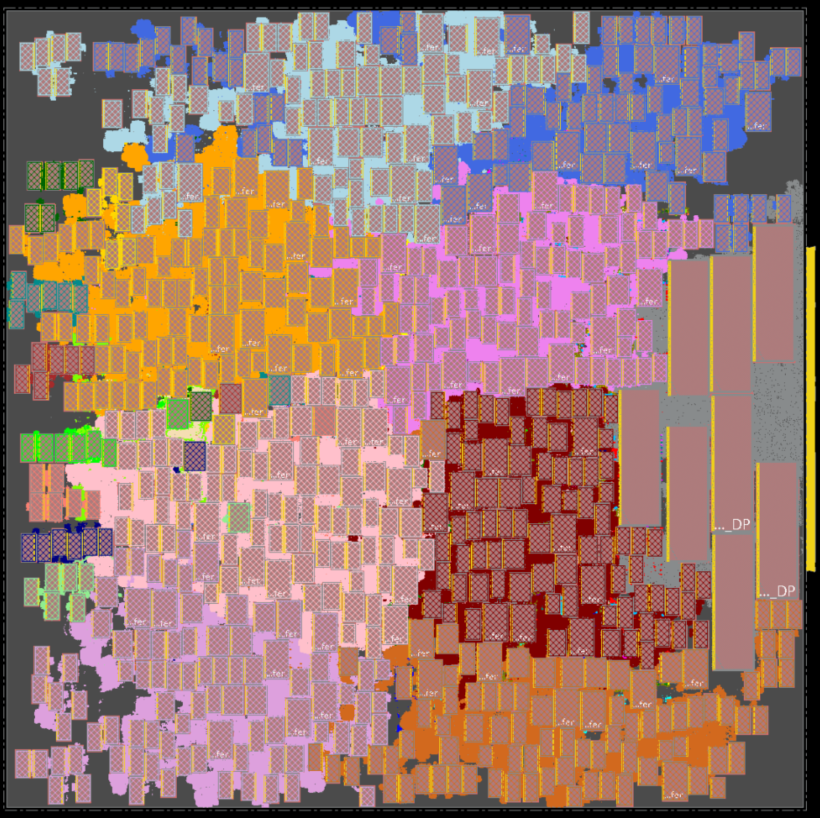}
        \caption{}
    \end{subfigure}
     \begin{subfigure}[b]{0.23\textwidth}
        \centering
        \includegraphics[width=\textwidth]{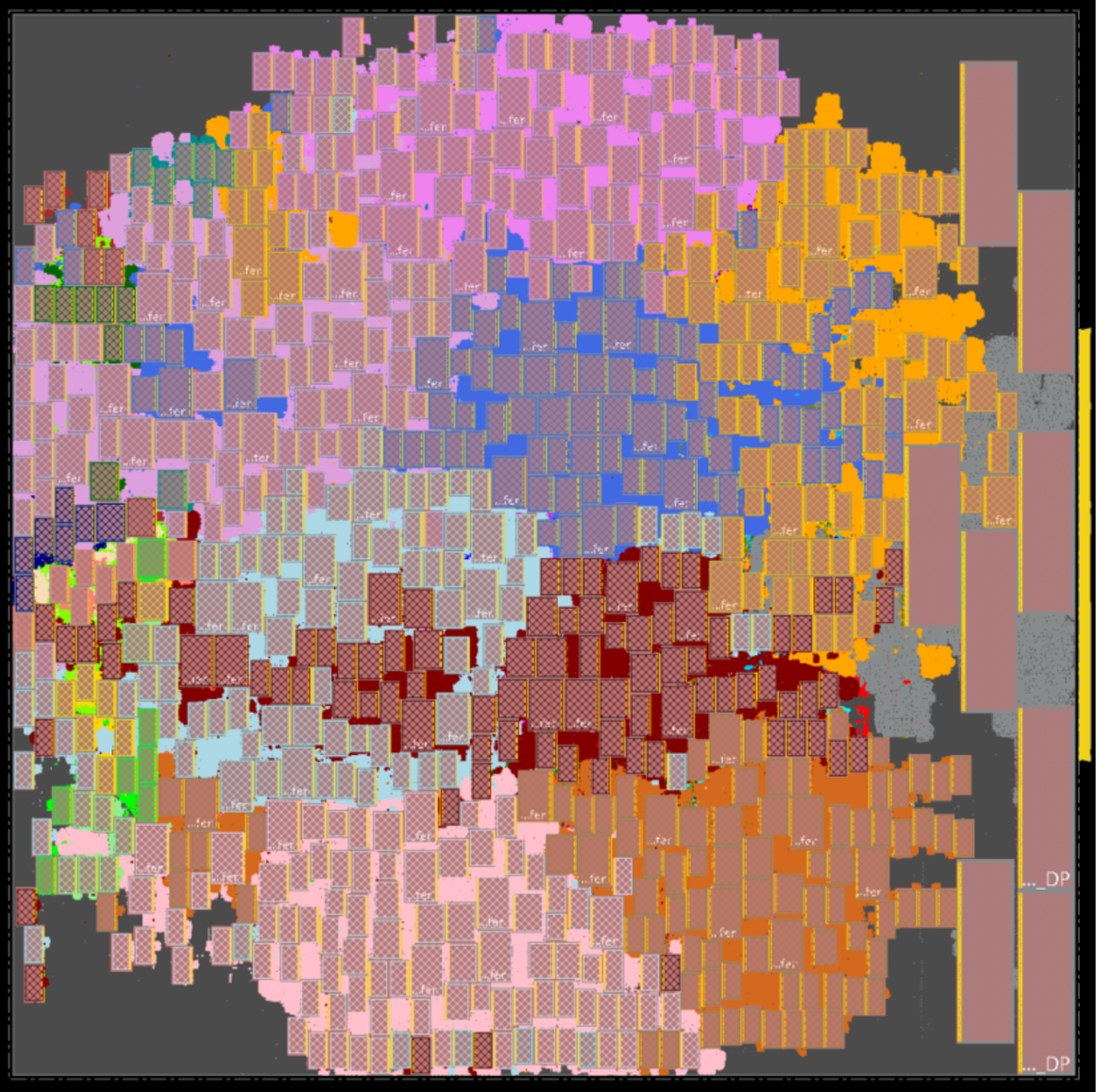}
        \caption{}
    \end{subfigure}
     \begin{subfigure}[b]{0.23\textwidth}
        \centering
        \includegraphics[width=\textwidth]{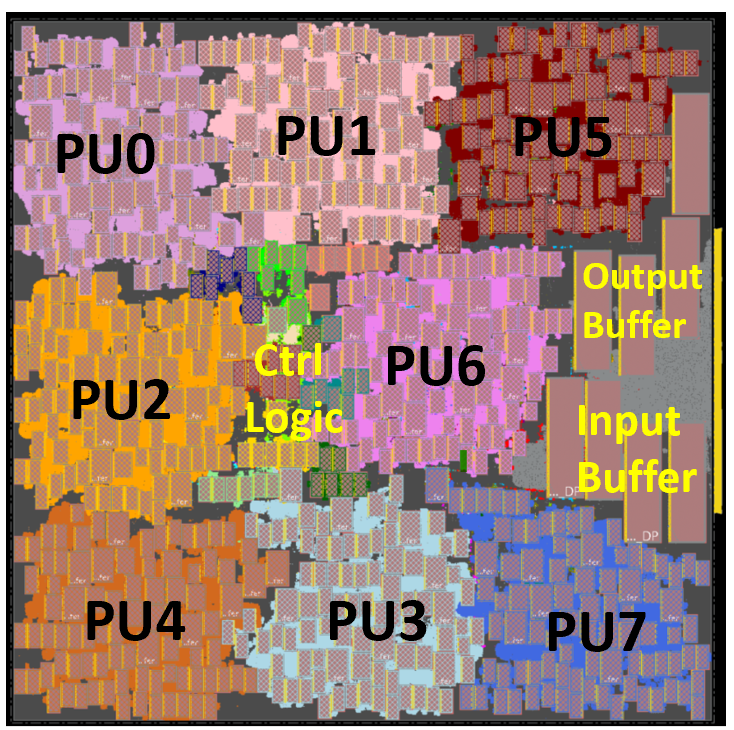}
        \caption{}
    \end{subfigure}
     \begin{subfigure}[b]{0.23\textwidth}
        \centering
        \includegraphics[width=\textwidth]{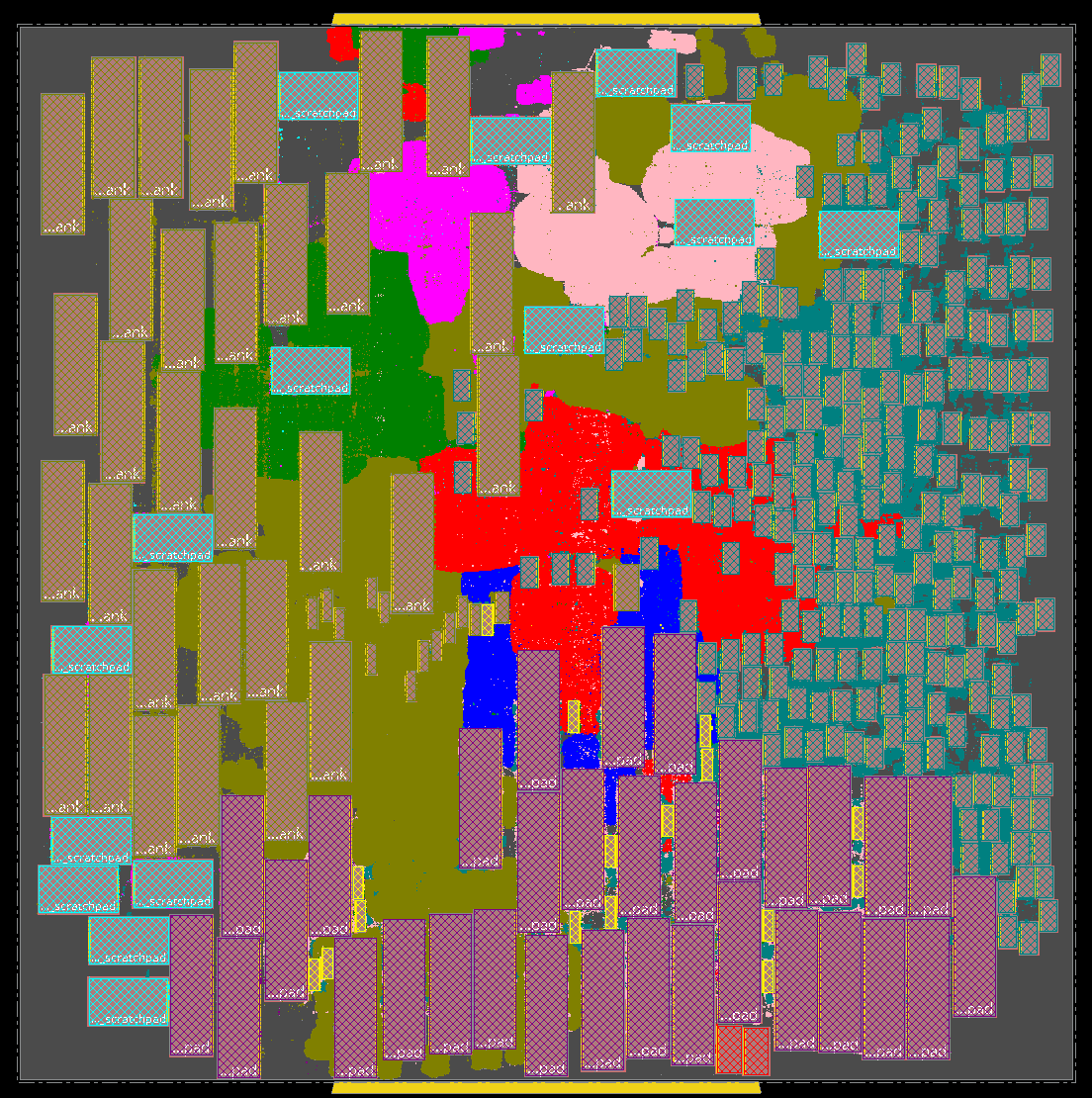}
        \caption{}
    \end{subfigure}
     \begin{subfigure}[b]{0.23\textwidth}
        \centering
        \includegraphics[width=\textwidth]{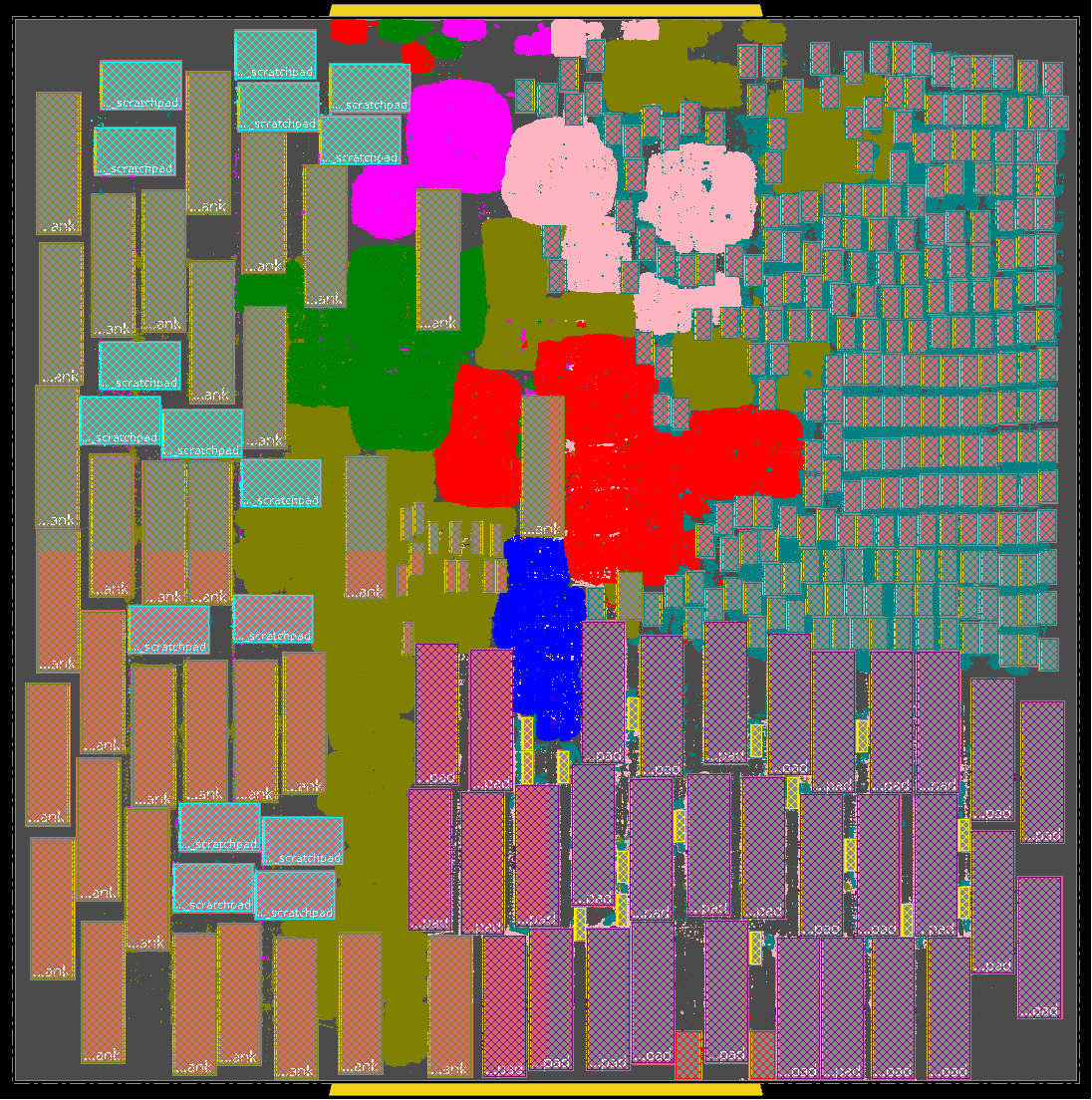}
        \caption{}
    \end{subfigure}
     \begin{subfigure}[b]{0.23\textwidth}
        \centering
        \includegraphics[width=\textwidth]{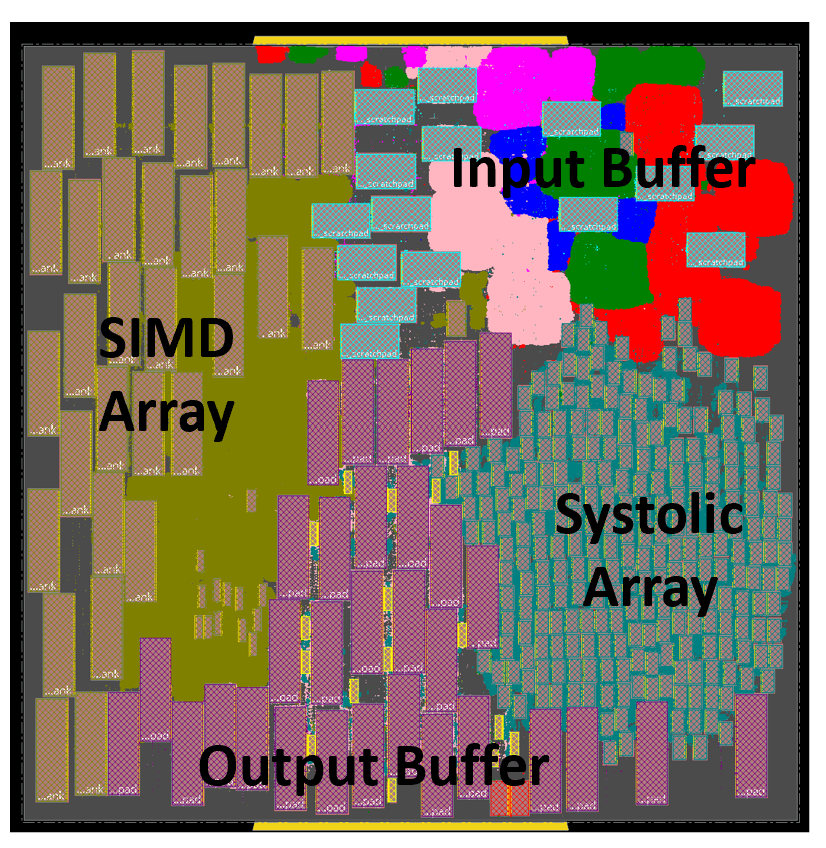}
        \caption{}
    \end{subfigure}
    \caption{Post-route layouts of Tabla3 and GeneSys02 designs with different flows. Design (Flow):
    (a) Tabla03 ({\em RePlAce});
    (b) Tabla03 ({\em DREAMPlace});
    (c) Tabla03 ({\em DG-RePlAce});
    (d) GeneSys02 ({\em RePlAce});
    (e) GeneSys02 ({\em DREAMPlace});
    (f) GeneSys02 ({\em DG-RePlAce}).
    For the same design, each module maintains consistent coloring across different layouts.}
    \vspace{-0.5cm}
    \label{fig:layout}
\end{figure}

We can observe \textcolor{black}{the} following conclusions.
\begin{itemize}
    \item Our approach outperforms both {\em RePlAce} and {\em DREAMPlace} in terms of routed wirelength, \textcolor{black}{achieving average reductions of $10\%$ and $7\%$,
    respectively.}
    
    \item Our approach outperforms both {\em RePlAce} and {\em DREAMPlace} in terms of TNS, \textcolor{black}{achieving average reductions of $31\%$ and $34\%$,
    respectively.}

    \item Our approach achieves similar speedup as {\em DREAMPlace} in terms of total turnaround time, but our approach is about $1.75X$ faster than {\em DREAMPlace} in terms of global placement runtime.
    The detailed runtime analysis is presented in Section \ref{subsec:runtime}.

    \item For the Tabla03 design, our approach significantly outperforms both {\em RePlAce} and {\em DREAMPlace} in all the metrics (wirelength, power and timing).
    We attribute this to {\em DG-RePlAce}'s ability to 
    identify the dataflow and datapath of the design, which enables it to generate the placement in accordance with dataflow and datapath constraints. 
    Figure \ref{fig:layout}(c) shows the post-route layout of the Tabla03 design for the global placement generated by {\em DG-RePlAce}. We can see that it perfectly matches the dataflow pattern illustrated in Figure \ref{fig:Tabla_dataflow}.
    \textcolor{black}{By contrast, in the layout from {\em DREAMPlace} (Figure \ref{fig:layout}(b)),
    we can see that PU2 (in orange) gets mixed up with other Processing Units (PUs), 
    leading to a significant degradation in wirelength, power and performance.
    {\em DREAMPlace}'s lack of awareness of the design's dataflow
    and datapath information means that it has more difficulty
    in generating placements that align with
    the dataflow and datapath structure of the design.}

    \item For the GeneSys02 design, our approach delivers significantly better timing
    compared to both {\em RePlAce} and {\em DREAMPlace}. 
    This is because {\em DG-RePlAce} follows the dataflow pattern inherent in the GeneSys02 design, as shown in Figure \ref{fig:layout}(f).  
    We also observe that the {\em Input Buffer} module (highlighted in purple) becomes mixed up with other modules when placed by {\em RePlAce} (Figure \ref{fig:layout}(d)) and {\em DREAMPlace} (Figure \ref{fig:layout}(e)).  
    While {\em DREAMPlace}'s solution  generates significantly better wirelength,
    its power improvement is very limited. We attribute this to the GeneSys02 design's
    extreme macro dominance, where the leakage power and \textcolor{black}{internal power (i.e., the power consumed by the CMOS circuit during the brief period when both PMOS and NMOS transistors are simultaneously switching as the logic changes its state \cite{WesteH10})} 
    constitute 
    67\% of total power consumption. We leave how to achieve a better power and 
    timing tradeoff as a direction for future work.


\end{itemize}

\subsection{Runtime Comparison Against {\em DREAMPlace}}
\label{subsec:runtime}

\textcolor{black}{We now} compare the runtime of {\em DG-RePlAce} against 
that of the leading
GPU-accelerated global placer, {\em DREAMPlace}.
As shown in Table \ref{tab:result},
the global placement runtime of {\em DG-RePlAce}
is less than that of {\em DREAMPlace},
while its overall turnaround time is similar.
We will first discuss the global placement runtime, 
and then examine the overall turnaround time.

\begin{figure}[!t]
    \centering
    \includegraphics[width=0.42\textwidth]{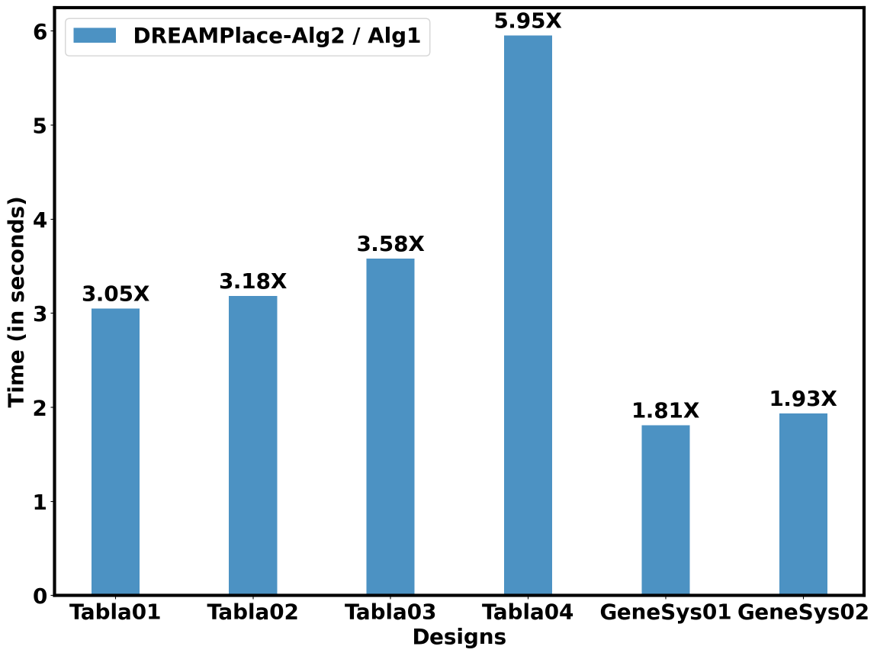}
    \caption{Runtime comparison for different implementations of wirelength gradient computation.}
    \label{fig:wl_runtime}
\end{figure}

The global placement runtime efficiency of {\em DG-RePlAce} can be attributed 
to the following two factors.
\begin{itemize}
    \item  {\em DG-RePlAce} achieves convergence with fewer iterations, due to an improved initial placement generated by our {\em Dataflow-Driven Initial Global Placement}.
    The iterations required for convergence of {\em RePlAce}, {\em DREAMPlace} and {\em DG-RePlAce} are presented in Table \ref{tab:iter}.\footnote{The {\em stop\_overflow} hyperparameter is 0.1 for {\em RePlAce}~\cite{RePlAce_params}, {\em DREAMPlace}~\cite{DREAMPlace_params} and {\em DG-RePlAce}.}  On average, {\em DG-RePlAce} achieves convergence in 24\% fewer iterations
    compared to {\em DREAMPlace}.

    \item Our parallel wirelength gradient algorithm (Algorithm \ref{alg:wirelength_grad})
outperforms the one used by {\em DREAMPlace}, \textcolor{black}{denoted} as {\em DREAMPlace-Alg2}.
For a fair comparison, we implement the wirelength gradient algorithm used by DREAMPlace 
(Algorithm 2 in \cite{LinJGLD20}). The result is shown in Figure \ref{fig:wl_runtime},
which suggests that our algorithm is 
on average $3.25X$ faster.
To further confirm that the increased runtime overhead of {\em DREAMPlace-Alg2} is due 
to high-fanout nets, we remove all nets connecting over 100 instances.\footnote{$ignore\_net\_degree$ is 100 by default in {\em DREAMPlace} \cite{DREAMPlace_params}.}
After removing all of these high-fanout nets,
{\em DREAMPlace-Alg2} achieves the same runtime as our Algorithm \ref{alg:wirelength_grad}.
\textcolor{black}{
This suggests that our  parallel wirelength gradient algorithm is effective on high-fanout nets.}

\end{itemize}

\begin{table}[!t]
    \caption{\textcolor{black}{Iterations} required for convergence of {\em RePlAce}, {\em DREAMPlace} and {\em DG-RePlAce}.
    We highlight best values in blue bold font.}
    \label{tab:iter}
    \centering
    \begin{tabular}{|c|c|c|c|}
    \hline 
    Design & {\em RePlAce} & {\em DREAMPlace} & {\em DG-RePlAce} \\ \hline
    Tabla01 & 410 & 450 & \textbf{\textcolor{blue}{380}} \\ \hline
    Tabla02 & 460 & 546 & \textbf{\textcolor{blue}{390}} \\ \hline
    Tabla03 & 460 & 507 & \textbf{\textcolor{blue}{380}} \\ \hline
    Tabla04 & 520 & 687 & \textbf{\textcolor{blue}{490}} \\ \hline
    GeneSys01 & 520 & 593 & \textbf{\textcolor{blue}{470}} \\ \hline
    GeneSys02 & 510 & 598 & \textbf{\textcolor{blue}{450}} \\ \hline
    \end{tabular}
    \end{table}

 \begin{figure}[!t]
    \centering
    \includegraphics[width=0.44\textwidth]{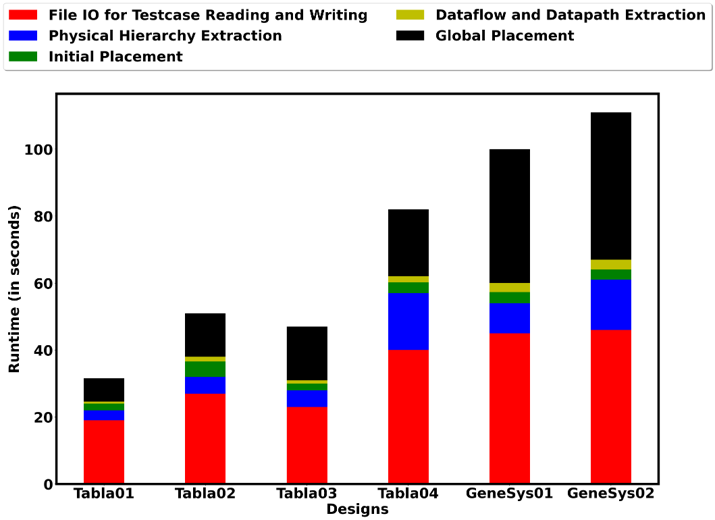}
    \caption{Runtime breakdown of {\em DG-RePlAce}.}
    \label{fig:runtime}
\end{figure}

The longer overall turnaround time for {\em DG-RePlAce}
results from the {\em file IO for testcase reading and writing}
and {\em physical hierarchy extraction}.
Figure \ref{fig:runtime} shows the
detailed runtime breakdown of {\em DG-RePlAce}.
We see that {\em file IO for testcase reading and writing} accounts for $50\%$ of 
the overall turnaround time. 
This is due to complexity of the industry-strength database 
(OpenDB \cite{OpenDB} in OpenROAD \cite{OpenROAD}) that we use, which brings
increased loading times for designs.
On the other hand, considering that the design is typically loaded just once,
substituting {\em DG-RePlAce} for {\em DREAMPlace} in scenarios where global
placement is executed many times (e.g., \textcolor{black}{
1,000 placement samples per design are obtained by 
{\em AutoDMP} \cite{AutoDMP}}) will significantly improve runtimes for such scenarios.


\begin{figure}[!t]
    \centering
    \begin{subfigure}[b]{0.22\textwidth}
        \centering
        \includegraphics[width=\textwidth]{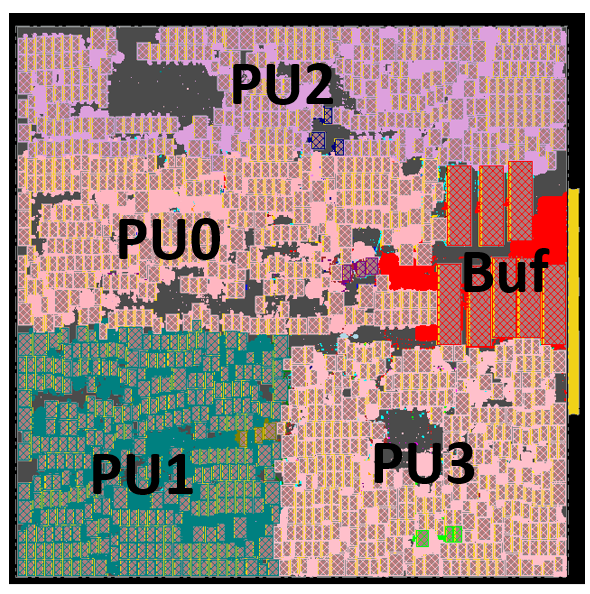}
        \caption{}
    \end{subfigure}
     \begin{subfigure}[b]{0.225\textwidth}
        \centering
        \includegraphics[width=\textwidth]{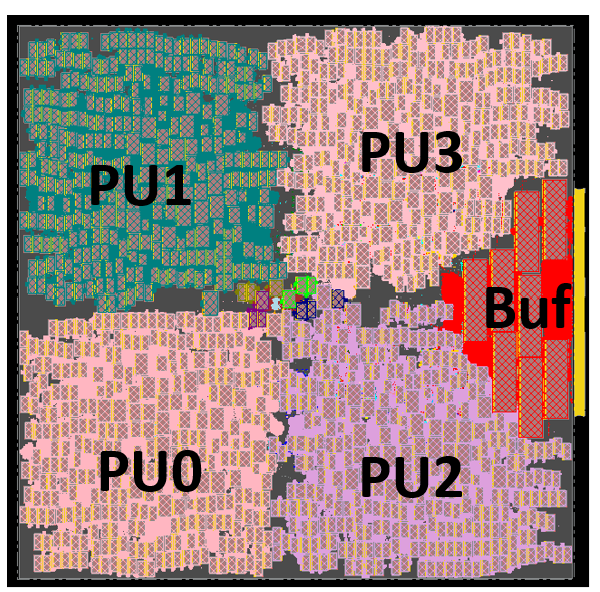}
        \caption{}
    \end{subfigure}
     \begin{subfigure}[b]{0.22\textwidth}
        \centering
        \includegraphics[width=\textwidth]{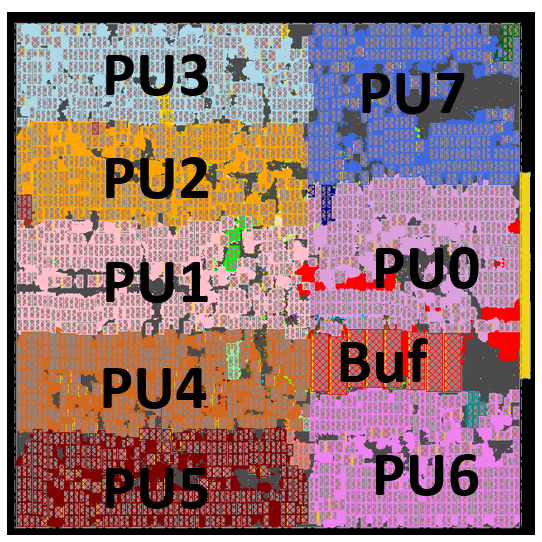}
        \caption{}
    \end{subfigure}
     \begin{subfigure}[b]{0.22\textwidth}
        \centering
        \includegraphics[width=\textwidth]{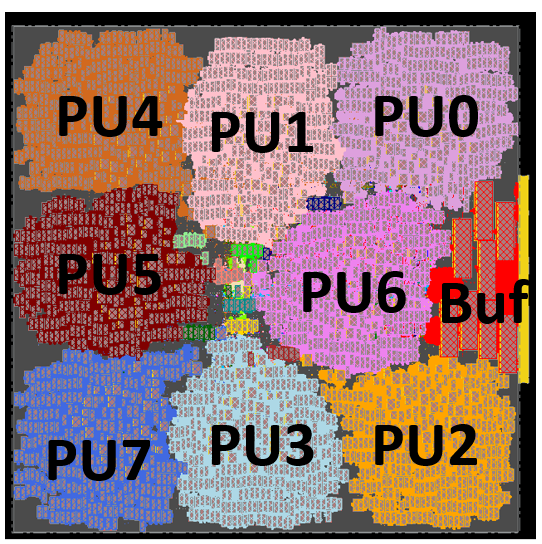}
        \caption{}
    \end{subfigure}
     \begin{subfigure}[b]{0.22\textwidth}
        \centering
        \includegraphics[width=\textwidth]{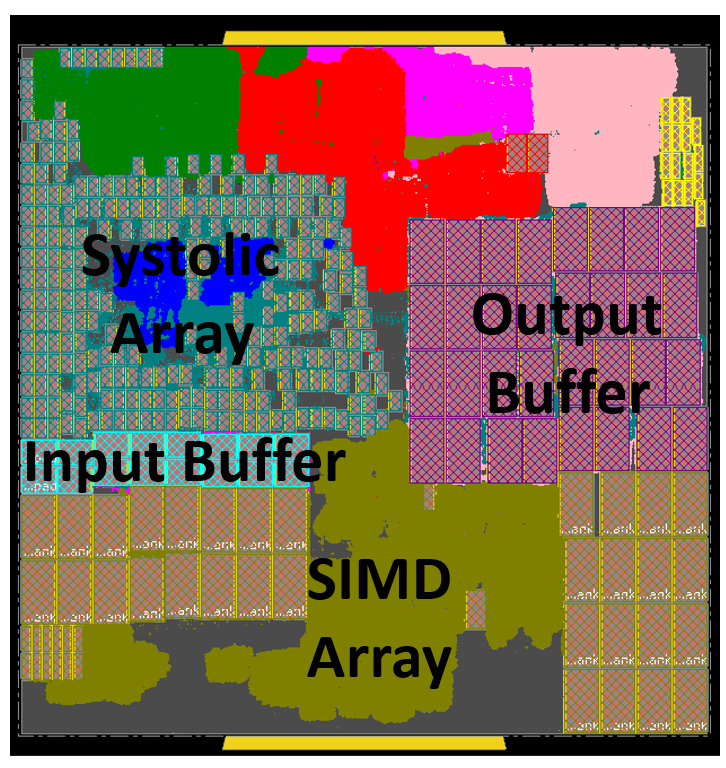}
        \caption{}
    \end{subfigure}
     \begin{subfigure}[b]{0.223\textwidth}
        \centering
        \includegraphics[width=\textwidth]{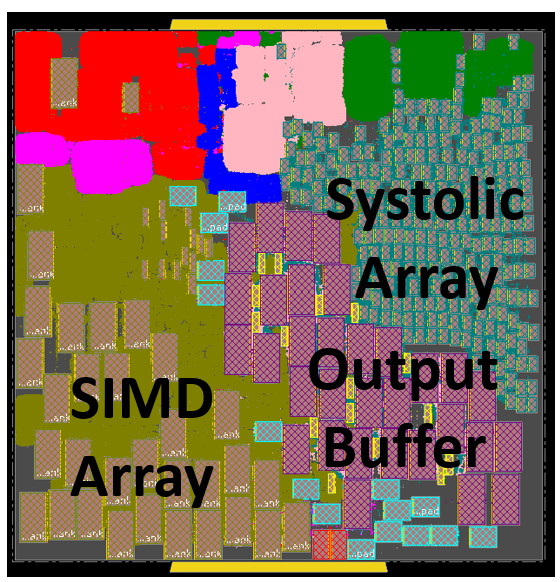}
        \caption{}
    \end{subfigure}
    \caption{Post-route layouts of Tabla02, Tabla04 and GeneSys01 designs with different flows. 
    (a) Tabla02 ({\em Hier-RTLMP});
    (b) Tabla02 ({\em DG-RePlAce});
    (c) Tabla04 ({\em Hier-RTLMP)};
    (d) Tabla04 ({\em DG-RePlAce});
    (e) GeneSys01 ({\em Hier-RTLMP)};
    (f) GeneSys01 ({\em DG-RePlAce}).
    For the same design, each module maintains consistent coloring across different layouts.
    In this figure,  ``Buf'' stands for input and output buffer.}
    \label{fig:hier_rtlmplayout}
\end{figure}

\subsection{Comparison with {Hier-RTLMP}}
\label{sec:comp_hier_rtlmp}

In this section, we compare {\em DG-RePlAce} with
the dataflow-driven multilevel
macro placer {\em Hier-RTLMP} \cite{KahngVW23}.
{\em Hier-RTLMP} uses the same physical hierarchy extraction approach and also considers dataflow information
when determining locations of macros.\footnote{We use the latest version of {\em Hier-RTLMP} from the OpenROAD repository \cite{Hier-RTLMP}.}
The results are presented in Table \ref{tab:result}, and Figure \ref{fig:hier_rtlmplayout} shows the post-route layouts.  According to Table \ref{tab:result},
{\em DG-RePlAce} achieves 15\% wirelength reduction compared 
to {\em Hier-RTLMP}. 
{\em Hier-RTLMP} has worse wirelength
because it models each cluster as a rectangular shape, as shown in Figure \ref{fig:hier_rtlmplayout}, potentially leading to unnecessary signal net detours.

\begin{figure}[!t]
    \centering
    \includegraphics[width=0.4\textwidth]{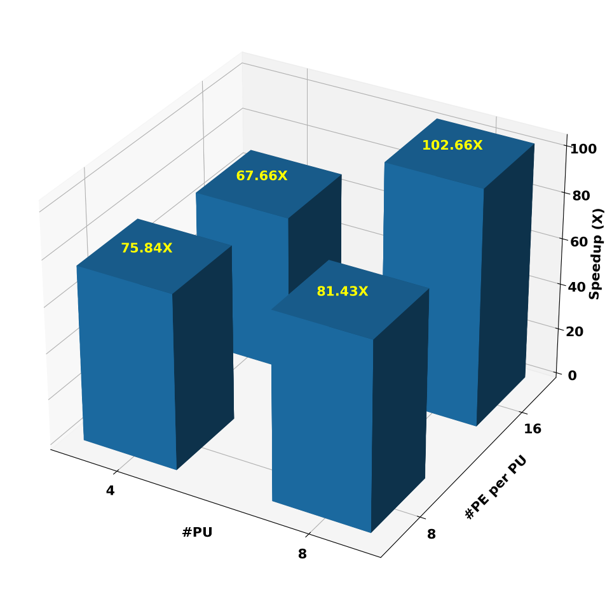}
    \caption{Speedup of {\em DG-RePlAce} over {\em Hier-RTLMP} for Tabla designs.}
    \label{fig:hier_rtlmp_runtime}
\end{figure}

Moreover, {\em Hier-RTLMP} fails to generate macro placement
for the GeneSys02 design. {\em Hier-RTLMP}
uses the Sequence Pair \cite{MurataFNK96} representation and Simulated Annealing \cite{KirkpatrickGV83} algorithm to determine shapes and locations for clusters level by level. Therefore, it may not be able to obtain a feasible solution when it tries to place macros within a cluster whose location and shape have been determined in the previous step.
Additionally, as has been pointed in \cite{AutoDMP},
the use of Simulated Annealing algorithm in {\em Hier-RTLMP}
makes it suffer from poor runtime scalability.
Figure \ref{fig:hier_rtlmp_runtime} shows how the speedup achieved by {\em DG-RePlAce} 
over {\em Hier-RTLMP} changes with the number of PUs (\#PU) and the number of 
PEs per PU (\#PE per PU) for Tabla designs.
We see that when the total number of PEs increases from
32 (Tabla01) to 128 (Tabla04), the speedup provided by 
{\em DG-RePlAce} over {\em Hier-RTLMP} increases from 76$X$ to 103$X$. Such speedups
are enabling for architects or front-end designers who seek to identify the 
optimal \#PU and \#PE per PU during the initial stages of machine learning 
accelerator development.

We also observe that for the GeneSys01 design,
{\em Hier-RTLMP} generates the best timing metrics in terms of both WNS 
and TNS.  We attribute this to the relatively low macro utilization of GeneSys01 
(see Table \ref{tab:benchmark}).
In such contexts, {\em Hier-RTLMP} is able to generate
reasonable macro tilings that are aligned with the dataflow structure, 
as shown in Figure \ref{fig:hier_rtlmplayout}(e).

\begin{table}[!t]
\caption{Effect of dataflow and datapath constraints (averages over all testcases).
We highlight best values of metrics in blue bold font.
Data points are normalized.
}
\label{tab:ablation_result}
\resizebox{0.99\columnwidth}{!} {
\centering
\begin{tabular}{|c|c|c|c|c|}
\hline 
Metrics & {\em RePlAce} & {\em DG-RePlAce$_{nf}$}
        & {\em DG-RePlAce$_{np}$} & {\em DG-RePlAce} \\ \hline
WL$_{avg}$  & 1.00 & 0.92 & 0.91 & \textbf{\textcolor{blue}{0.90}} \\ \hline
TNS$_{avg}$ & 1.00 & 0.61 & 0.80 & \textbf{\textcolor{blue}{0.61}}      \\ \hline
\end{tabular}
}
\end{table}

\subsection{Ablation Study}
\label{sec:ablation}

To demonstrate the effect of dataflow and datapath constraints, 
we run an ablation study \cite{ablation} by removing dataflow or datapath 
constraints to understand their respective contributions to the overall 
performance of {\em DG-RePlAce}.
We conduct two separate experiments using variants of {\em DG-RePlAce}. 
The first variant, referred to as {\em DG-RePlAce$_{nf}$}, is
executed without the dataflow constraint.
The second variant, designated as {\em DG-RePlAce$_{np}$}, 
is executed without the datapath constraint.
These modifications allow us to assess the individual contributions of each constraint. The experimental results are presented in Table \ref{tab:ablation_result}.
In this table, WL$_{avg}$ and TNS$_{avg}$ respectively represent the average normalized routed wirelength and average total negative slack over all the testcases in Table \ref{tab:benchmark}, 
compared with those from {\em RePlAce}.
We observe that both {\em DG-RePlAce$_{nf}$} and {\em DG-RePlAce$_{np}$}
generate better results than {\em RePlAce} in terms of TNS, but {\em DG-RePlAce} always generates the best results.
This suggests that both dataflow and datapath constraints are 
important components of {\em DG-RePlAce}.

\subsection{Results on TILOS MacroPlacement Benchmarks}
\label{sec:tilos_result}

\begin{table}[!b]
\caption{Experimental results on {\em TILOS MacroPlacement} benchmarks.
We highlight best values of metrics in blue bold font.
Data points for WL, Power, WNS and TNS are normalized.
{\em DREAMPlace*} represents running {\em DREAMPlace} with updated hyperparameters: $ignore\_net\_threshold$ = 1e9 and $iterations$ = 5000.
}
\label{tab:result_tilos}
\resizebox{1\columnwidth}{!} {
\centering
\begin{tabular}{|c|c|c|c|c|c|c|c|}
\hline 
\multicolumn{1}{|l|}{\makecell{Design}}             
    & \makecell{Global Placer}
    & \makecell{WL}   
    & \makecell{Power}
    & \makecell{ WNS} 
    & \makecell{TNS}
    & \makecell{GP \\  ($s$) }
    & \makecell{ TAT\\($s$)}
    \\ \hline \Xhline{2\arrayrulewidth}
\multirow{3}{*}{\makecell{BlackParrot}}
& {\em RePlAce} & 1.00  & 1.00  & -0.123  & -108.15  & 387  & 653  \\  \cline{2-8}
& {\em DREAMPlace} & 0.92  & 0.98  & -0.023  & -2.623  & 61  
                   & \textbf{\textcolor{blue}{88}}  \\  \cline{2-8}
& {\em DG-RePlAce} & \textbf{\textcolor{blue}{0.90}}  
                   & \textbf{\textcolor{blue}{0.97}} 
                   & \textbf{\textcolor{blue}{-0.014}}  
                   & \textbf{\textcolor{blue}{-0.078}}  
                   & \textbf{\textcolor{blue}{32}}  
                   & 200  \\  \cline{2-8}
\hline  \Xhline{2\arrayrulewidth}
\multirow{4}{*}{\makecell{MemPool \\ Group}}
& {\em RePlAce} & 1.00  & 1.00  & -0.073  & -99.989  & 1896  & 2712  \\  \cline{2-8}
& {\em DREAMPlace} & \textbf{\textcolor{blue}{0.92}} 
                   & \textbf{\textcolor{blue}{0.97}}  
                   & -0.086  & -134.421  
                   & \textbf{\textcolor{blue}{72}} 
                   & \textbf{\textcolor{blue}{167}}  \\  \cline{2-8}
& {\em DREAMPlace*} & \textbf{\textcolor{blue}{0.92}}
                    & \textbf{\textcolor{blue}{0.97}}
                    & -0.069
                    & -108.193
                    & 178 
                    & 284 \\ \cline{2-8}
& {\em DG-RePlAce} & 0.95  & 0.98  
                   & \textbf{\textcolor{blue}{-0.067}}  
                   & \textbf{\textcolor{blue}{-38.71}}  & 122  & 591  \\  \cline{2-8}
\hline
\end{tabular}
}
\end{table}

\begin{figure}
    \centering
    \begin{subfigure}[b]{0.23\textwidth}
        \centering
        \includegraphics[width=\textwidth]{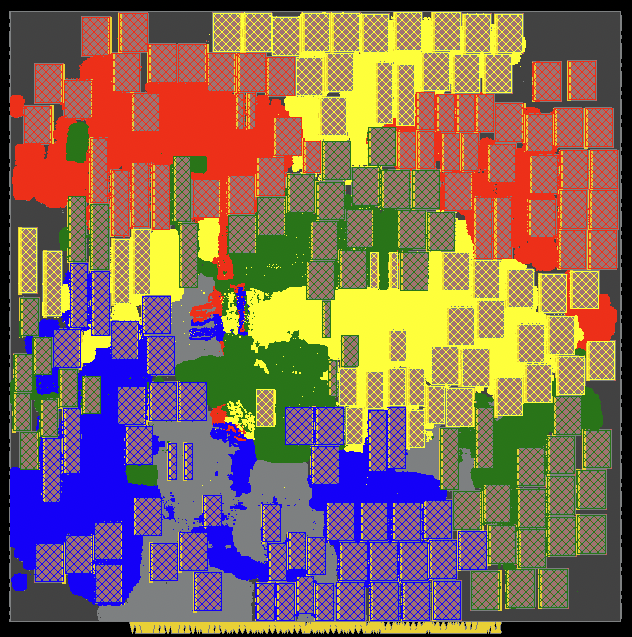}
        \caption{}
    \end{subfigure}
     \begin{subfigure}[b]{0.23\textwidth}
        \centering
        \includegraphics[width=0.99\textwidth]{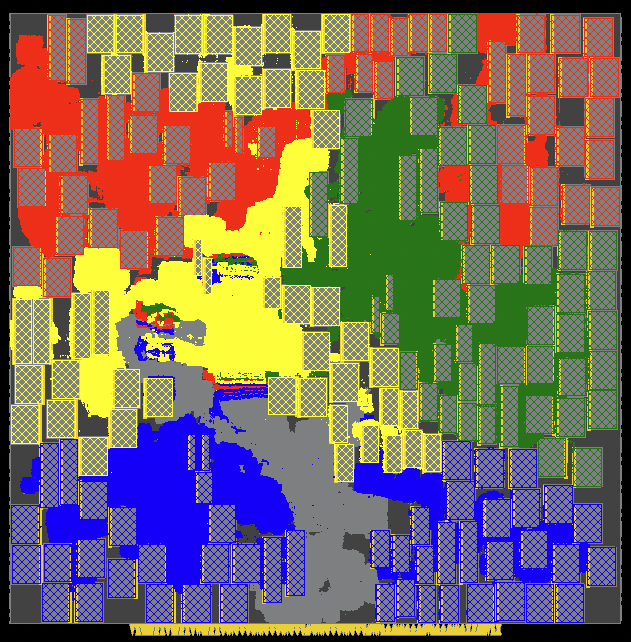}
        \caption{}
    \end{subfigure}
     \begin{subfigure}[b]{0.23\textwidth}
        \centering
        \includegraphics[width=\textwidth]{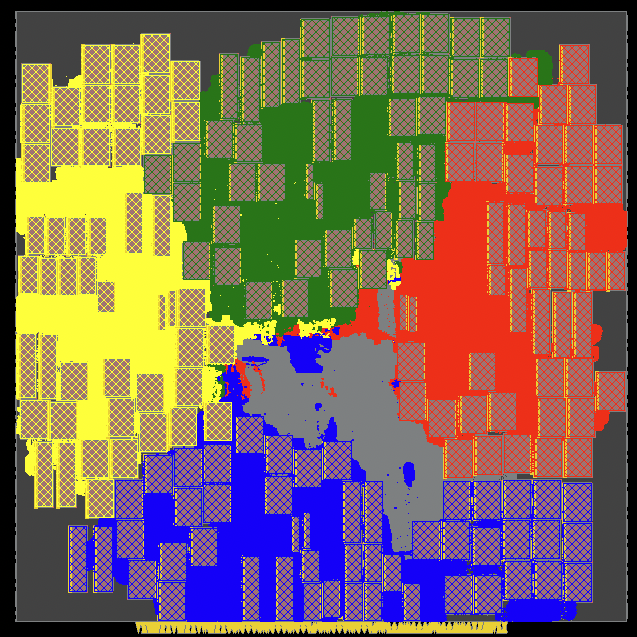}
        \caption{}
    \end{subfigure}
     \begin{subfigure}[b]{0.23\textwidth}
        \centering
        \includegraphics[width=\textwidth]{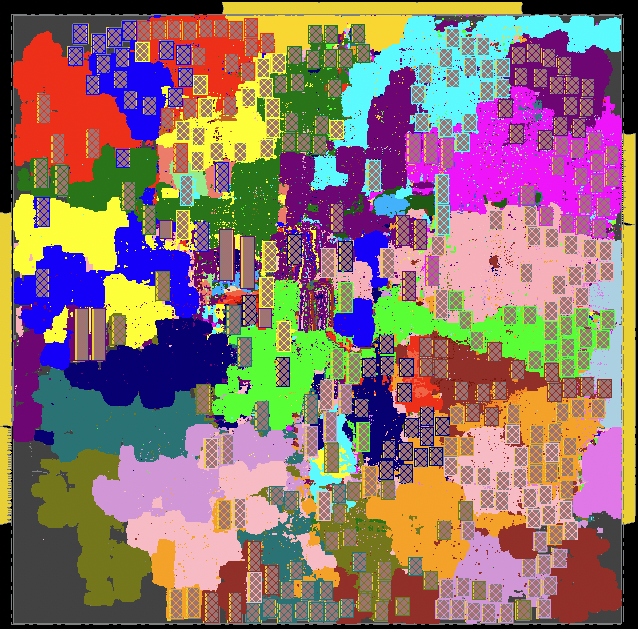}
        \caption{}
    \end{subfigure}
     \begin{subfigure}[b]{0.23\textwidth}
        \centering
        \includegraphics[width=\textwidth]{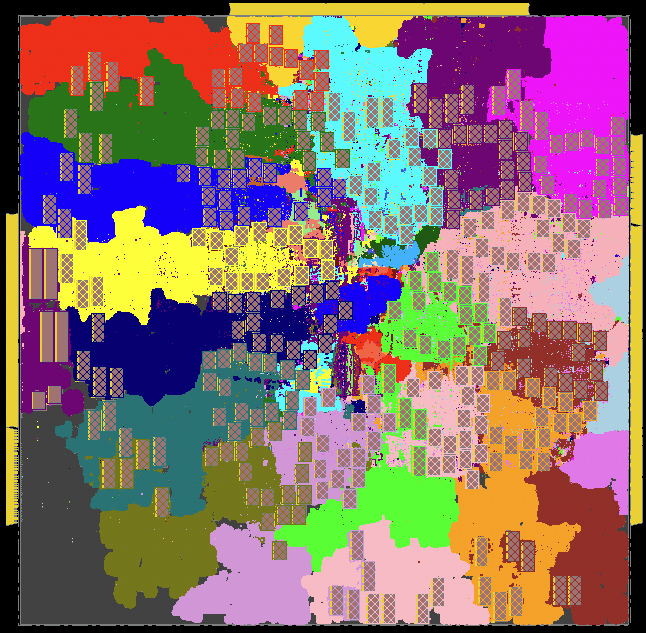}
        \caption{}
    \end{subfigure}
     \begin{subfigure}[b]{0.23\textwidth}
        \centering
        \includegraphics[width=\textwidth]{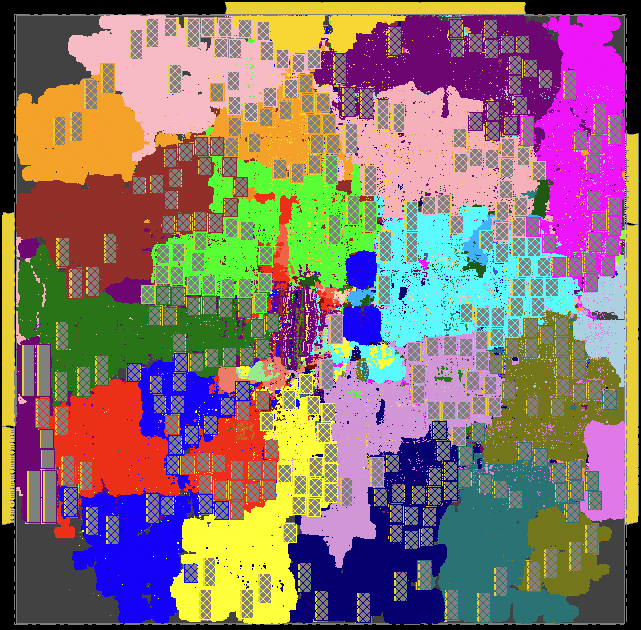}
        \caption{}
    \end{subfigure}
    \caption{Post-route layouts of BlackParrot and MemPool Group designs with different flows.
    Design (Flow):
    (a) BlackParrot ({\em RePlAce});
    (b) BlackParrot ({\em DREAMPlace});
    (c) BlackParrot ({\em DG-RePlAce});
    (d) MemPool Group ({\em RePlAce});
    (e) MemPool Group ({\em DREAMPlace});
    (f) MemPool Group ({\em DG-RePlAce}).
    \textcolor{black}{The layouts from {\em DREAMPlace} are generated with the default hyperparameter settings \cite{DREAMPlace_params}.}
    For the same design, each module maintains consistent coloring across different layouts.}
    \label{fig:tilos_layout}
\end{figure}

The dataflow-driven approach proposed in this work is inspired by the 
unique demands of dataflow and datapath structures in modern, highly scaled
machine learning accelerators.
However, benefits of our proposed method may reach beyond machine learning 
accelerators.
To demonstrate the generality and effectiveness of our {\em DG-RePlAce},
we conduct evaluations on the two largest benchmarks -- 
BlackParrot (Quad-Core)~\cite{BlackParrot} (827K instances, 196 macros) 
and MemPool Group~\cite{MemPool} (2529K instances, 326 macros) --
from the {\em TILOS MacroPlacement Benchmarks} \cite{MacroPlacement}. 
The experimental results are summarized in Table \ref{tab:result_tilos},
and the post-route layouts are presented in Figure \ref{fig:tilos_layout}.

For the BlackParrot design, {\em DG-RePlAce} dominates {\em RePlAce} and {\em DREAMPlace} across all metrics, including wirelength, power and timing.
    Figures \ref{fig:tilos_layout}(a), (b) and (c) show the post-route layouts from 
    {\em RePlAce}, {\em DREAMPlace} and {\em DG-RePlAce}, respectively. 
    It is clear that
    one of the CPU cores (marked in yellow) gets mixed up when 
    placed by {\em RePlAce} and {\em DREAMPlace}, resulting in 
    worse wirelength, power and timing.
    Additionally, we notice that {\em DG-RePlAce} is 2$X$ faster 
    than {\em DREAMPlace} in terms of global placement runtime, but 
    has total turnaround time larger than {\em DREAMPlace}. This 
    is because it takes 134 seconds to load the design into OpenROAD.
    Subtracting the loading time, the turnaround time for
    {\em DG-RePlAce} drops to 66 seconds. 
    These findings are consistent with the runtime analysis presented in Section \ref{subsec:runtime}.

For the MemPool Group design, {\em DG-RePlAce} achieves 
significantly better timing (TNS)
compared to both {\em RePlAce} and {\em DREAMPlace}.
However, {\em DG-RePlAce} suffers from 3\% wirelength degradation over {\em DREAMPlace}.
To understand this wirelength increase, we 
use early global route (eGR) in a commercial place-and-route 
tool (Cadence Innovus 21.1) to examine the congestion 
maps for placements generated by {\em DG-RePlAce} and {\em RePlAce}. 
We observe that
the placement from {\em DG-RePlAce} is free from congestion,
while there are 0.05\% horizontal and 0.02\% vertical congestion in the placement
from {\em DREAMPlace}. 
\textcolor{black}{
This accounts for the increased wirelength with {\em DG-RePlAce},
as {\em DG-RePlAce} tries to distribute instances more evenly\
due to the bloat-shrink methodology (see Section \ref{sec:initial_placement}) -- but at the cost of wirelength degradation.} 
We leave how to achieve better wirelength and congestion tradeoffs
as a direction for future work.

We also notice that {\em DREAMPlace}, using its default parameter settings \cite{DREAMPlace_params},
\textcolor{black}{terminates because it reaches its maximum number 
of iterations ($iteration$ = 1000 by default).}
To prevent such early termination, we rerun {\em DREAMPlace} 
with updated hyperparameters that leave ample margin:
$ignore\_net\_threshold$ = 1e9 and $iterations$ = 5000; we denote
these runs as {\em DREAMPlace*}.
These hyperparameters are also set to be the default limits on net
filtering and iteration count for {\em RePlAce} and {\em DG-RePlAce}.
As shown in Table \ref{tab:result_tilos},
{\em DREAMPlace*} delivers better results compared to the 
default configuration of {\em DREAMPlace},
but at the cost of increased global placement runtime and 
total turnaround time.
Even with the updated hyperparameters, {\em DG-RePlAce} continues to outperform {\em DREAMPlace*} in terms of timing metrics (WNS and TNS).

\begin{table} [!t]
\caption{Runtime breakdown for the MemPool Group and
MegaBoom\_X4 benchmarks.
Effective TAT is the net turnaround time, calculated 
by subtracting the time spent on handling input and output files 
(IO) from the total turnaround time (TAT).
{\em DREAMPlace*} represents running {\em DREAMPlace} with 
the updated hyperparameters $ignore\_net\_threshold$ = 1e9 and $iterations$ = 5000.
}
\label{tab:runtime_mg}
\resizebox{1\columnwidth}{!} {
\centering
\begin{tabular}{|c|c|c|c|c|c|c|}
\hline 
\multicolumn{1}{|l|}{\makecell{Design}}       
    & \makecell{Global \\ Placer} 
    & \makecell{Convergence \\ Iterations} 
    & \makecell{GP \\  ($s$)}
    & \makecell{IO \\  ($s$) }
    & \makecell{ TAT \\ ($s$)}
    & \makecell{ Effective \\ TAT ($s$)}
    \\ \hline \Xhline{2\arrayrulewidth}
\multirow{4}{*}{\makecell{MemPool \\ Group}}
& {\em RePlAce}  & 690 & 1896 & 332  & 2712 & 2380 \\  \cline{2-7}
& {\em DREAMPlace} & 1001
                   & 72
                   & 95
                   & 167 
                   & 72 \\  \cline{2-7}
& {\em DREAMPlace*} & 1091
                    & 178
                    & 95
                    & 284
                    & 189 \\ \cline{2-7}
& {\em DG-RePlAce} & 530 & 122 &  332 & 591 & 259 \\  \cline{2-7} \Xhline{2\arrayrulewidth}
\multirow{4}{*}{\makecell{MegaBoom\_X4}}
& {\em RePlAce}  & 870 & 7433 & 370 & 8546 &  8176 \\  \cline{2-7}
& {\em DREAMPlace} & 993
                   & 319
                   & 230
                   & 550
                   & 319 \\  \cline{2-7}
& {\em DREAMPlace*} & 1036
                    & 881
                    & 230
                    & 1113
                    & 883  \\ \cline{2-7}
& {\em DG-RePlAce} & 770 & 418 &  370 & 937 & 567 \\  \cline{2-7}
\hline
\end{tabular}
}
\end{table}

\begin{table} [!t]
\caption{Experimental results on the MegaBoom\_X4 design.
Data points for WL are normalized.
}
\label{tab:result_megaboom}
\resizebox{1\columnwidth}{!} {
\centering
\begin{tabular}{|c|c|c|c|c|c|c|}
\hline 
\multicolumn{1}{|l|}{\makecell{Design}}       
    & \makecell{\# Std \\  Cells} 
    & \makecell{\# Nets} 
    & \makecell{Global \\ Placer} 
    & \makecell{WL}
    & \makecell{Horizontal \\ Congestion}
    & \makecell{Vertical \\ Congestion}
    \\ \hline \Xhline{2\arrayrulewidth}
\multirow{4}{*}{\makecell{MegaBoom\_X4}} &
\multirow{4}{*}{\makecell{5807K}} &
\multirow{4}{*}{\makecell{5831K}} 
& {\em RePlAce}  & 1.00 & 0.01\% & 0.07\% \\  \cline{4-7}
& & & {\em DREAMPlace} & 1.00 & 0.02\% & 0.08\%  \\  \cline{4-7}
& & &  {\em DREAMPlace*} & 1.00
                    & 0.01\%
                    & 0.08\%  \\ \cline{4-7}
& & &  {\em DG-RePlAce} & 1.00 & 0.00\% &  0.08\%  \\  \cline{4-7}
\hline
\end{tabular}
}
\end{table}

\begin{figure}[!h]
    \centering
    \begin{subfigure}[b]{0.35\textwidth}
        \centering
        \includegraphics[width=\textwidth]{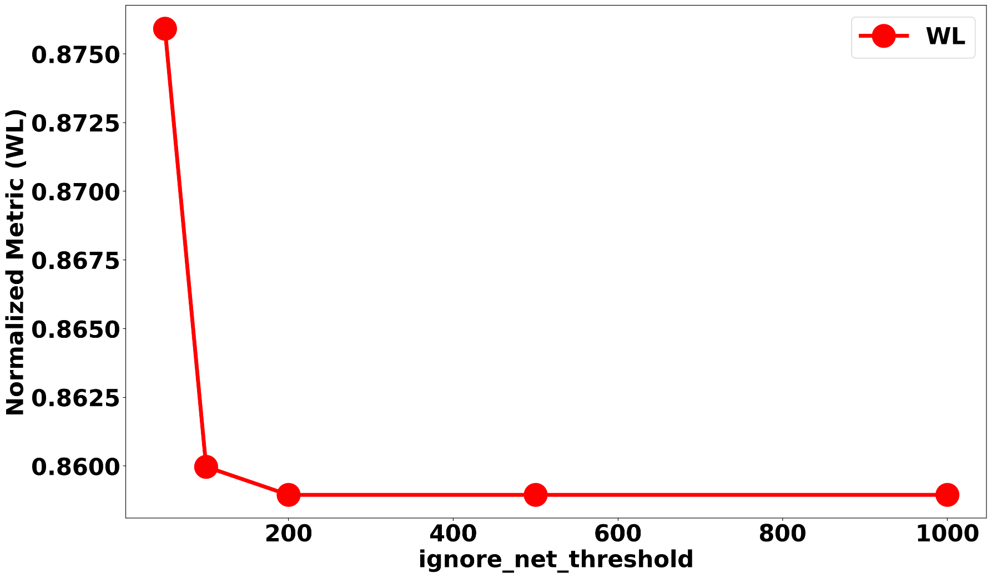}
        \caption{}
    \end{subfigure}
     \begin{subfigure}[b]{0.35\textwidth}
        \centering
        \includegraphics[width=\textwidth]{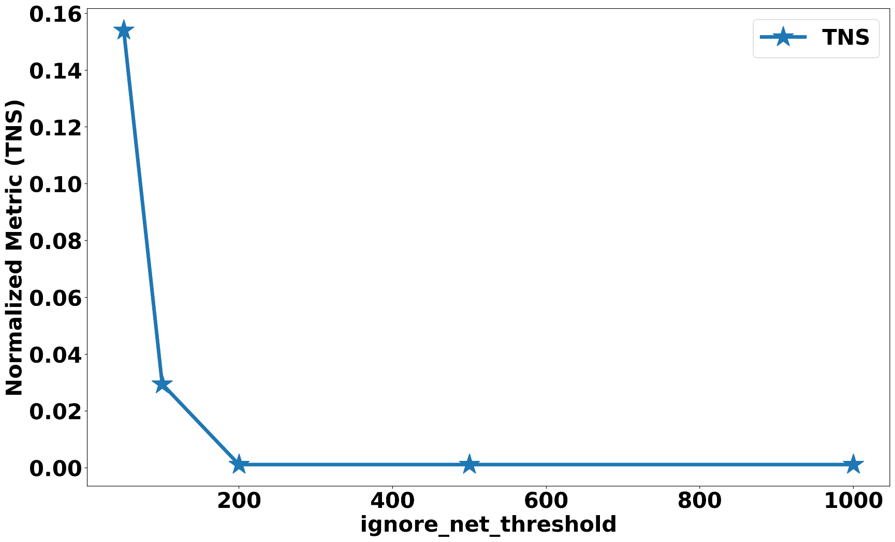}
        \caption{}
    \end{subfigure}
     \begin{subfigure}[b]{0.35\textwidth}
        \centering
        \includegraphics[width=\textwidth]{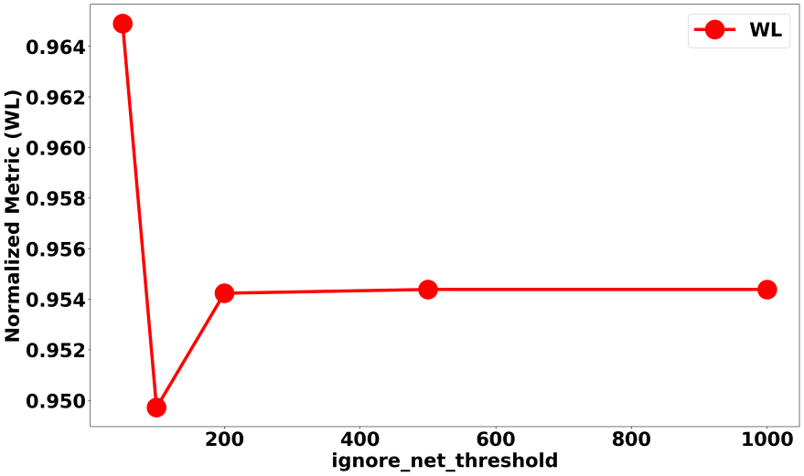}
        \caption{}
    \end{subfigure}
     \begin{subfigure}[b]{0.35\textwidth}
        \centering
        \includegraphics[width=\textwidth]{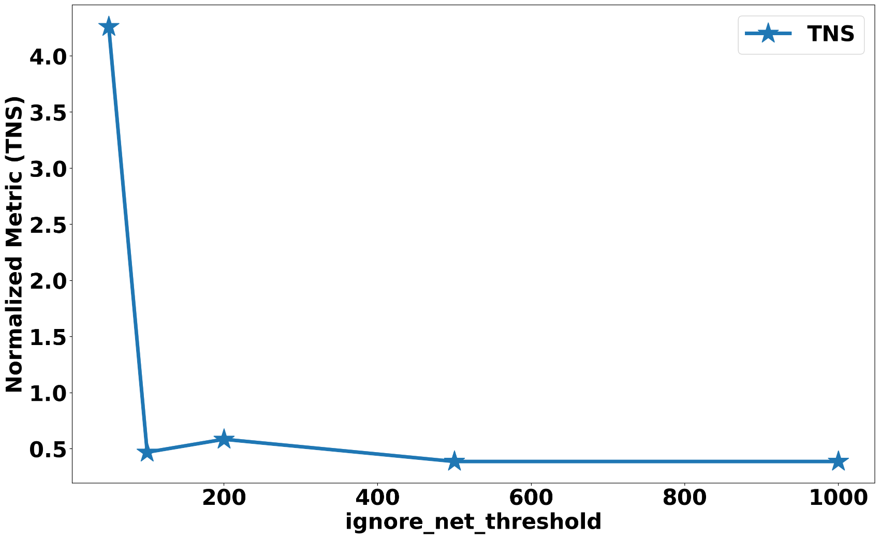}
        \caption{}
    \end{subfigure}
    \caption{\textcolor{black}{Effect of $ignore\_net\_threshold$ on {\em DG-RePlAce}. All the metrics are collected after completion
    of post-route optimization and normalized as described in 
    Section \ref{subsec:main_results}.
    (a) BlackParrot (normalized wirelength);
    (b) BlackParrot (normalized TNS);
    (c) MemPool Group (normalized wirelength);
    and (d) MemPool Group (normalized TNS).}}
    \label{fig:net_threshold}
\end{figure}

Additionally, for the MemPool Group design, the global placement runtime of {\em DG-RePlAce}
exceeds that of {\em DREAMPlace}. To delve into the reasons behind the global placement runtime degradation, we examine the detailed runtime breakdown for both {\em DG-RePlAce} and {\em DREAMPlace}.
The results are presented in Table \ref{tab:runtime_mg}. 
We can observe the following conclusions.
\begin{itemize}
    \item The global placement runtime of {\em DREAMPlace} increases 
    by 2.47$\times$ when considering all the signal nets during placement.
    \item With the same hyperparameter settings, {\em DG-RePlAce} 
    is 1.46$\times$ faster than {\em DREAMPlace*} in terms of global 
    placement runtime, while its total turnaround time is larger 
    than that of {\em DREAMPlace*}.

    \item {\em DG-RePlAce} converges in fewer iterations compared to 
    {\em DREAMPlace*},  which accounts for its 1.46$X$ speedup 
    of global placement runtime. \textcolor{black}{The runtime per iteration of {\em DG-RePlAce} is longer than that of {\em DREAMPlace*},
    due to the additional pseudo nets introduced by the dataflow constraints (Section \ref{sec:initial_placement})
    and datapath constraints (Section \ref{sec:dataflow_extraction}).
    By removing these pseudo nets, the global placement runtime decreases from 122 seconds to 68 seconds.} 
    To further confirm this speedup,
    we run the same experiments on another large design,
    MegaBoom\_X4 (four-core RISC-V MegaBoom \cite{BOOM}), 
    which has more than 5.8M instances.
    Experimental results on MegaBoom\_X4 \textcolor{black}{(shown in 
    Table \ref{tab:runtime_mg})} give support to our analysis.
    Detailed metrics for MegaBoom\_X4 are presented in Table 
    \ref{tab:result_megaboom}.\footnote{Since we cannot finish 
    the place-and-route flow for the MegaBoom\_X4 design, we
    report metrics (wirelength, horizontal and vertical congestion)
    after early global routing in the commercial tool.}

    \item \textcolor{black}{We also study the effect of $ignore\_net\_threshold$ on {\em DG-RePlAce}. We sweep the $ignore\_net\_threshold$ values across 50, 100, 200, 500, 1000.
    The results are shown in Figure \ref{fig:net_threshold}.
    We can see that increasing $ignore\_net\_threshold$ from 50 to 1000 results in approximately $1\%$ improvement in wirelength.
    However, the improvement in TNS is significantly more substantial.}

    \item After subtracting the time spent on handling file input 
    and output (IO) from the total turnaround time (TAT), the 
    net turnaround time (``Effective TAT'') of {\em DG-RePlAce} is  
    \textcolor{black}{smaller} than that of  {\em DREAMPlace*}. 
    As pointed out in Section \ref{subsec:runtime}, we attribute
    this to the {\em physical hierarchy extraction} process.  
    Enhancing the efficiency of the {\em physical hierarchy extraction}
    process is a key objective for our future research efforts.  
\end{itemize}

\section{Conclusion and Future work}
\label{sec:conclusion}

In this work, we develop {\em DG-RePlAce},
a new and fast GPU accelerated global placement framework which is
built on top of the OpenROAD infrastructure \cite{OpenROAD}, and
which exploits the inherent dataflow and datapath structures of 
machine learning accelerators to achieve superior results.
Experimental results show that {\em DG-RePlAce} outperforms both 
{\em RePlAce} and {\em DREAMPlace} in terms of routed wirelength 
and total negative slack metrics.
Extensions to {\em DG-RePlAce} that we are currently exploring 
include:
(i) incorporation of density screens
for routability and virtual resizing for timing optimization;
(ii) application of ML-based multi-objective optimization methods 
to autotune the hyperparameters of {\em DG-RePlAce}, 
potentially achieving better tradeoffs across wirelength, 
congestion, power and timing;
and (iii) improving the runtime of the {\em physical hierarchy extraction} process.
In combination with open-sourcing and OpenROAD integration, 
we believe \textcolor{black}{that} this work will add to the
foundations for new \textcolor{black}{research} on fast 
and high-quality global placers.



\begin{IEEEbiography}
[{\raisebox{0.3in}{\includegraphics[width=1in, height=1in, clip, keepaspectratio]
{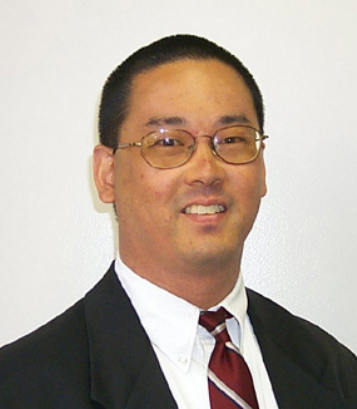}}}]
{Andrew B. Kahng} is a distinguished professor in the CSE and ECE Departments of the University of California at San Diego. His interests include IC physical design, the design-manufacturing interface, large-scale combinatorial optimization, AI/ML for EDA and IC design, and technology roadmapping. He received the Ph.D. degree in Computer Science from the University of California at San Diego.
\end{IEEEbiography}

\vspace{-6.5cm}

\begin{IEEEbiography}
[{\raisebox{0.4in}{\includegraphics[width=0.9in, height=1in, clip,keepaspectratio]
{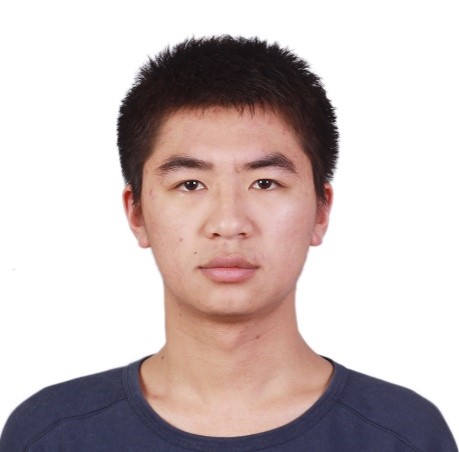}}}]
{Zhiang Wang}
received the M.S. degree in electrical and computer
engineering from the University of California at San Diego, La Jolla, 
in 2022. He is currently pursuing the Ph.D. degree at the University of California 
at San Diego, La Jolla. His current research interests include 
partitioning, placement methodology and optimization.
\end{IEEEbiography}

\clearpage
\newpage

\end{document}